\documentclass[iop,floatfix]{emulateapj}

\pdfoutput=1

\RequirePackage{color}
\RequirePackage{code}
\input{astro.sty}

\usepackage[T1]{fontenc}% (2) specify encoding

\newcommand{\asinh}{\mathop{\rm asinh}\nolimits}

\shorttitle{Pixel Processing in PS1}
\shortauthors{C.Z. Waters et al}
\begin{document}
\title{Pan-STARRS Pixel Processing : Detrending, Warping, Stacking}

% this is a crude trick to get the order of affiliations right.  These
% names are used in the affiliations below.  The user needs to (1) set
% the order and numbers to have the correct sequence in the author
% list and (2) re-order the list at the bottom (and comment-out as needed)
\def\IfA{1}
\def\Princeton{2}
\def\DUR{3}
\def\STSCI{4}
\def\Pitt{5}
%\def\CfA{2}
%\def\MPIA{3}
%\def\USNO{4}
%\def\JHU{1}

% This example has a first author from UH:
\author{
C.~Z. Waters,\altaffilmark{\IfA}
E.~A. Magnier,\altaffilmark{\IfA}
P.~A. Price,\altaffilmark{\Princeton}
K.~C. Chambers,\altaffilmark{\IfA} 
W.~S. Burgett,\altaffilmark{\IfA}
P. Draper,\altaffilmark{\DUR}
H.~A. Flewelling,\altaffilmark{\IfA}
K. W. Hodapp,\altaffilmark{\IfA}
M.~E. Huber,\altaffilmark{\IfA}
R. Jedicke,\altaffilmark{\IfA}
N. Kaiser,\altaffilmark{\IfA}
R.-P. Kudritzki,\altaffilmark{\IfA}
R.~H. Lupton,\altaffilmark{\Princeton}
 N. Metcalfe,\altaffilmark{\DUR}
A. Rest,\altaffilmark{\STSCI}
W.~E. Sweeney,\altaffilmark{\IfA}
J.~L. Tonry, \altaffilmark{\IfA}
R.~J. Wainscoat,\altaffilmark{\IfA} 
W.~M. Wood-Vasey\altaffilmark{\Pitt}
}
%PS1 Builders
%PS Builder List

% L. Denneau,\altaffilmark{\IfA}
% T. Grav,\altaffilmark{\IfA}
% J. N. Heasley,\altaffilmark{\IfA}
% G. A. Luppino,\altaffilmark{\IfA}
% D. G. Monet,\altaffilmark{\USNO}
% J.~S. Morgan,\altaffilmark{\IfA}
% P. M. Onaka,\altaffilmark{\IfA}
% C.~W. Stubbs,\altaffilmark{\CfA}
 % this bracket terminates author list

% The ordering here should be sequential, matching the sequence in the list of authors:
\altaffiltext{\IfA}{Institute for Astronomy, University of Hawaii, 2680 Woodlawn Drive, Honolulu HI 96822}
% \altaffiltext{\CfA}{Harvard-Smithsonian Center for Astrophysics, 60 Garden Street, Cambridge, MA 02138}
\altaffiltext{\Princeton}{Department of Astrophysical Sciences, Princeton University, Princeton, NJ 08544, USA}
\altaffiltext{\STSCI}{Space Telescope Science Institute, 3700 San Martin Drive, Baltimore, MD 21218, USA}
\altaffiltext{\DUR}{Department of Physics, Durham University, South Road, Durham DH1 3LE, UK}
\altaffiltext{\Pitt}{Pittsburgh Particle Physics, Astrophysics, and Cosmology Center (PITT PACC), Physics and Astronomy Department, University of Pittsburgh, Pittsburgh, PA 15260, USA}
% \altaffiltext{\USNO}{US Naval Observatory, Flagstaff Station, Flagstaff, AZ 86001, USA}
% \altaffiltext{\JHU}{Department of Physics and Astronomy, Johns Hopkins University, 3400 North Charles Street, Baltimore, MD 21218, USA}
% \altaffiltext{\MPIA}{Max Planck Institute for Astronomy, K\"onigstuhl 17, D-69117 Heidelberg, Germany}
\begin{abstract}

The Pan-STARRS1 Science Consortium has carried out a set of imaging
surveys using the 1.4 gigapixel GPC1 camera on the PS1 telescope.  As
this camera is composed of many individual electronic readouts, and
covers a very large field of view, great care was taken to ensure that
the many instrumental effects were corrected to produce the most
uniform detector response possible.  We present the image detrending
steps used as part of the processing of the data contained within the
public release of the Pan-STARRS1 Data Release 1 (DR1).  In addition
to the single image processing, the methods used to transform the
375,573 individual exposures into a common sky-oriented grid are
discussed, as well as those used to produce both the image stack and
difference combination products.

\end{abstract}

% insert additional keywords as appropriate:

%% NOTE 2018.12.06 EAM : Things that still need to be done prior to submission:
%% * generate a valid bibtex entry for the diff paper, include in refs
%% * better covariance discussion (wait until analysis.tex version written)
%% * annotate the ghost, glint, and perhaps other images with arrows to illustrate features
%% * check on total size vs arxiv limits

\section{Introduction}

The 1.8m Pan-STARRS\,1 telescope is located on the summit of Haleakala
on the Hawaiian island of Maui.  The wide-field optical design of the
telescope \citep{2004SPIE.5489..667H} produces a 3.3 degree field of view with
low distortion and minimal vignetting even at the edges of the
illuminated region.  The optics and natural seeing combine to yield
good image quality: 75\% of the images have full-width half-max values
less than (1.51, 1.39, 1.34, 1.27, 1.21) arcseconds for (\grizy), with
a floor of $\sim 0.7$ arcseconds.

The \PSONE\ camera \citep{2009amos.confE..40T}, known as GPC1, consists of a
mosaic of 60 back-illuminated CCDs manufactured by Lincoln Laboratory.
The CCDs each consist of an $8\times8$ grid of $590 \times 598$
pixel readout regions, yielding an effective $4846 \times 4868$
detector.  Initial performance assessments are presented in
\cite{2008SPIE.7014E..0DO}.  Routine observations are conducted remotely from the
Advanced Technology Research Center in Kula, the main facility of the
University of Hawaii's Institute for Astronomy (IfA) operations on Maui.
The Pan-STARRS1 filters and photometric system have already been
described in detail in \cite{2012ApJ...750...99T}.

For nearly 4 years, from 2010 May through 2014 March, this telescope
was used to perform a collection of astronomical surveys under the
aegis of the Pan-STARRS Science Consortium.  The majority of the time
(56\%) was spent on surveying the $\frac{3}{4}$ of the sky north of
$-30$ Declination with \grizy\ filters in the so-called $3\pi$ Survey.
Another $\sim 25\%$ of the time was concentrated on repeated deep
observations of 10 specific fields in the Medium-Deep Survey.  The
rest of the time was used for several other surveys, including a
search for potentially hazardous asteroids in our solar system.  The
details of the telescope, surveys, and resulting science publications
are described by \cite{chambers2017}.  The Pan-STARRS1 filters and
photometric system has already been described in detail in
\cite{2012ApJ...750...99T}.

Pan-STARRS produced its first large-scale public data release, Data
Release 1 (DR1) on 16 December 2016.  DR1 contains the results of the
third full reduction of the Pan-STARRS $3\pi$ Survey archival data,
identified as PV3.  Previous reductions \citep[PV0, PV1, PV2;
  see][]{magnier2017.datasystem} were used internally for pipeline
optimization and the development of the initial photometric and
astrometric reference catalog \citep{magnier2017.calibration}.  The
products from these reductions were not publicly released, but have
been used to produce a wide range of scientific papers from the
Pan-STARRS 1 Science Consortium members \citep{chambers2017}.  DR1
contained only average information resulting from the many individual
images obtained by the $3\pi$ Survey observations.  A second data
release, DR2, was made available 28 January 2019.  DR2 provides
measurements from all of the individual exposures, and include an
improved calibration of the PV3 processing of that dataset.

This is the third in a series of seven papers describing the
Pan-STARRS1 Surveys, the data reduction techniques and the resulting
data products. This paper (Paper III) describes the details of the
pixel processing algorithms, including detrending, warping, adding (to
create stacked images), and subtracting (to create difference images),
along with the resulting image products and their properties.

%Chambers et al. 2017 (Paper I)
%The Pan-STARRS\,1 Surveys
\citet[][Paper I]{chambers2017} provide an overview of the Pan-STARRS
System, the design and execution of the Surveys, the resulting image
and catalog data products, a discussion of the overall data quality
and basic characteristics, and a brief summary of important results.

%Magnier et al. 2017 (Paper II)
%Pan-STARRS Data Processing Stages
\citet[][Paper II]{magnier2017.datasystem} describe how the various
data processing stages are organized and implemented in the Imaging
Processing Pipeline (IPP), including details of the the processing
database which is a critical element in the IPP infrastructure.

%Waters et al. 2017 (Paper III)
%Pan-STARRS Pixel Processing : Detrending, Warping, Stacking
%\citet[][Paper III]{waters2017}
% THIS PAPER

%Magnier et al. 2017 (Paper IV)
%Pan-STARRS Pixel Analysis : Source Detection
\citet[][Paper IV]{magnier2017.analysis} describe the details of the
source detection and photometry, including point-spread-function and
extended source fitting models, and the techniques for ``forced''
photometry measurements.

%Magnier et al. 2017 (Paper V)
%Pan-STARRS Photometric and Astrometric Calibration
\citet[][Paper V]{magnier2017.calibration} describe the final
calibration process, and the resulting photometric and astrometric
quality.

%Flewelling et al. 2017 (Paper VI)
%Pan-STARRS 1 Database and Data Products
\citet[][Paper VI]{flewelling2017}
describe the details of the resulting catalog data and its organization
in the Pan-STARRS database.

%Huber et al. 2017 (Paper VII)
\citet[][Paper VII]{huber2017} describe the Medium Deep Survey in
detail, including the unique issues and data products specific to that
survey. The Medium Deep Survey is not part of Data Releases 1 or 2 and
will be made available in a future data release.

% \note{DS notes fonts are not consistent for keywords, etc}

\section{Background}

The Pan-STARRS 1 Science Survey used the 1.4 gigapixel GPC1 camera
with the PS1 telescope on Haleakala Maui to image the sky north of
$-30^\circ$ declination.  The GPC1 camera is composed of 60 orthogonal
transfer array (OTA) devices arranged in an $8\times{}8$ grid,
excluding the four corners.  Each of the 60 devices is itself an
$8\times{}8$ grid of readout cells.  The large number of cells
parallelizes the readout process, reducing the overhead in each
exposure.  However, as a consequence, many calibration operations are
needed to ensure the response is consistent across the entire seven
square degree field of view.

The Pan-STARRS image processing pipeline (IPP) is described elsewhere
\citep{magnier2017.datasystem}, but a short summary follows.  The raw
image data is stored on the processing cluster, with a database
containing the metadata of exposure parameters.  These raw images can
be launched for the initial \IPPstage{chip} stage processing.  This
stage performs the image detrending (described below in section
\ref{sec:detrending}), as well as the single epoch photometry
\citep{magnier2017.analysis}, in parallel on the individual OTA device
data.  Following the \IPPstage{chip} stage is the \IPPstage{camera}
stage, in which the astrometry and photometry for the entire exposure
is calibrated by matching the detections against a reference catalog.
This stage also performs masking updates based on the now-known
positions and brightnesses of stars that create dynamic features (see
Section \ref{sec:dynamic_masks} below).  The \IPPstage{warp} stage is
the next to operate on the data, transforming the detector oriented
\IPPstage{chip} stage images onto common sky oriented images that have
fixed sky projections (Section \ref{sec:warping}).  When all
\IPPstage{warp} stage processing is done for a region of the sky,
\IPPstage{stack} processing is performed (Section \ref{sec:stacking})
to construct deeper, fully populated images from the set of
\IPPstage{warp} images that cover that region of the sky.  Transient
features are identified in the \IPPstage{diff} stage, which takes
input \IPPstage{warp} and/or \IPPstage{stack} data and performs image
differencing (Section \ref{sec:diffs}).  Further photometry is
performed in the \IPPstage{staticsky} and \IPPstage{skycal} stages,
which add extended source fitting to the point source photometry of
objects detected in the \IPPstage{stack} images, and again calibrate
the results against a reference catalog.  The \IPPstage{fullforce}
stage takes the catalog output of the \IPPstage{skycal} stage, and
uses the objects detected in that to perform forced photometry on the
individual \IPPstage{warp} stage images.  The details of these stages
are provided in \citet{magnier2017.analysis}.

A limited version of the same reduction procedure described above is also
performed in real time on new exposures as they are observed by the
telescope.  This process is automatic, with new exposures being
downloaded from the summit to the main IPP processing cluster at the
Maui Research and Technology Center in Kihei, and registered into the
processing database.  New \IPPstage{chip} stage reductions are
launched for science exposures, advancing processing upon completion
through to the \IPPstage{diff} stage, skipping the additional stack
and forced warp photometry stages.  This automatic processing allows
the ongoing solar system moving object search to identify candidates
for follow up observations within 24 hours of the initial set of
observations \citep{2015IAUGA..2251124W}.

Section \ref{sec:detrending} provides an overview of the detrending
process that corrects the instrumental signatures of GPC1, with
details of the construction of the reference detrend templates in
Section \ref{sec:detrend construction}.  An analysis of the algorithms
used to perform the \IPPstage{warp} (section \ref{sec:warping}),
\IPPstage{stack} (section \ref{sec:stacking}), and \IPPstage{diff}
(section \ref{sec:diffs}) stage transformations of the image data
follows after the list of detrend steps.  Finally, a discussion of the
remaining issues and possible future improvements is presented in
section \ref{sec:discussion}.

As mentioned above, the GPC1 camera is composed of 60 orthogonal
transfer array (OTA) devices arranged in an $8\times{}8$ grid,
excluding the four corners.  Each of the 60 devices is itself an
$8\times{}8$ grid of readout cells consisting of $590 \times 598$
pixels.  We label the OTAs by their coordinate in the camera grid in
the form `OTAXY', where X and Y each range from 0 - 7, e.g., OTA12 would
be the chip in the $(1,2)$ position of the grid.  Similarly, we
identify the cells as `xyXY' where X and Y again each range from 0 -
7.  

Image products presented in figures have been mosaicked to arrange
pixels as follows.  Single cell images are arranged such that pixel
$(1,1)$ is at the lower right corner (for example Figure
\ref{fig:burntool images}).  This corrects the parity difference
between the raw data and the sky.  Images mosaicked to show a full OTA
detector are arranged as they are on the focal plane (as in Figure
\ref{fig:dark image}.  The OTAs to the left of the midplane
(OTA4Y-OTA7Y) are oriented with cell xy00 and pixel $(590,1)$ to the
lower right of their position.  Due to the electronic connections of
the OTAs in the focal plane, the OTAs to the right of the midplane
(OTA0Y-OTA3Y) are rotated 180 degrees, and are oriented with cell xy00
and pixel $(590,1)$ to the top left of their position. For mosaics of
the full field of view, the OTAs are arranged as they see the sky,
with the cells arranged as in the single OTA images (Figure
\ref{fig:optical ghosts}).  The lower left corner is the empty
location where OTA70 would exist.  Toward the right, the OTA labels
decrease in $X$ label, with the empty OTA00 located in the lower
right.  The OTA $Y$ labels increase upward in the mosaic.

%%\textit{Note: These papers are being placed on the arXiv.org to
%%  provide crucial support information at the time of the public
%%  release of Data Release 1 (DR1).  We expect the arXiv versions to be
%%  updated prior to submission to the Astrophysical Journal in January
%%  2017.  Feedback and suggestions for additional information from early
%%  users of the data products are welcome during the submission and
%%  refereeing process.}

\section{GPC1 Detrend Details}
\label{sec:detrending}

Ensuring a consistent and uniform detector response across the
three-degree diameter field of view of the GPC1 camera is essential to
a well calibrated survey.  Many standard image detrending steps are
done for GPC1, with overscan subtraction removing the detector bias
level, dark frame subtraction to remove temperature and exposure time
dependent detector glows, and flat field correction to remove pixel to
pixel response functions.  We also perform fringe correction for the
reddest data in the \yps{} filter to remove the interference patterns
that arise in that filter due to the variations in the thickness of
the detector surface.

These corrections assume that the detector response is linear across
the full dynamic range and that the pixels contain only signals coming
from the imaged portion of the sky, or from linear dark current
sources within the detector.  This assumption is not universally true
for GPC1, and an additional set of detrending steps are required as a
result.  The first of these is the \IPPprog{burntool} correction,
which removes the flux trails left by the incomplete transfer of
charge along the readout columns.  These trails are generally only
evident for the brightest stars, as only pixels that are at or beyond
the saturation point of the detector leave residual charge.  A second
confounding effect is the non-linearity at the faint end of the pixel
range.  Some readout cells and some readout cell edge pixels
experience a sag relative to the linear trend at low illumination,
such that faint pixels appear fainter than expected.  The correction
to this requires amplifying the pixel values in these regions to match
the linear response.

Large regions of some OTA cells experience significant charge transfer
issues, making them unusable for science observations.  These regions
are therefore masked in processing, with these CTE regions making up
the largest fraction of masked pixels on the detector.  Other regions
are masked for reasons such as static bad pixel features or temporary
readout masking caused by issues in the camera electronics that make
these regions unreliable.  These all contribute to the detector mask,
a 16 bit value which records the reason a pixel is masked based on the
value added.  This mask is augmented in each exposure for dynamic
features that are masked based on the astronomical features within the
field of view.

Within the IPP, all detrending is done by the \IPPprog{ppImage}
program.  This program applies the detrend corrections to the
individual cells, and then an OTA-level mosaic is constructed for the
signal image, the mask image, and the variance map image.  The single
epoch photometry is done at this stage as well.  The following
subsections (\ref{sec:overscan} - \ref{sec:background}) detail the
detrending process used on GPC1 that are common to other detectors.
The GPC1 specific detrending steps are included after, explaining
these additional steps that remove the instrument signature.

\subsection{Overscan}
\label{sec:overscan}

Each cell on GPC1 has an overscan region that covers the first 34
columns of each row, and the last 10 rows of each column.  No light
lands on these pixels, so the science region is trimmed to exclude
them.  Each row has an overscan value subtracted, calculated by
finding the median value of that row's overscan pixels and then
smoothing between rows with a three-row boxcar median.

\subsection{Dark/Bias Subtraction}
\label{sec:dark}

\begin{figure}
  \centering
  \begin{minipage}{0.45\hsize}
    \includegraphics[width=0.9\hsize,angle=0,clip]{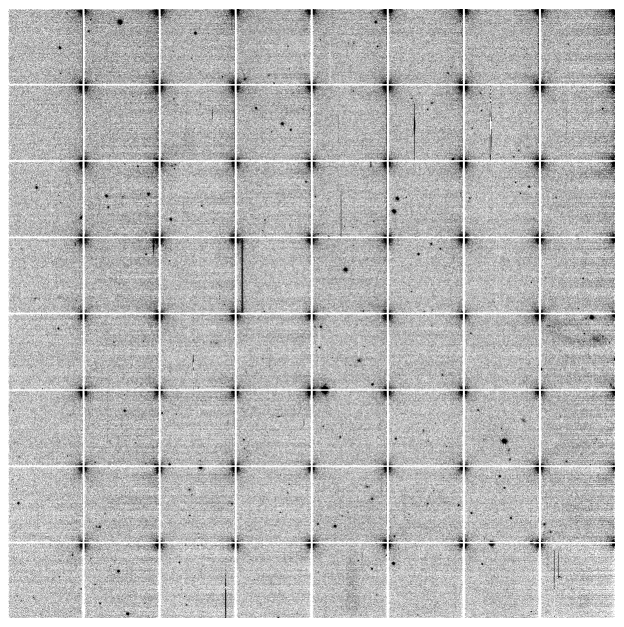}
  \end{minipage}%
  \begin{minipage}{0.45\hsize}
    \includegraphics[width=0.9\hsize,angle=0,clip]{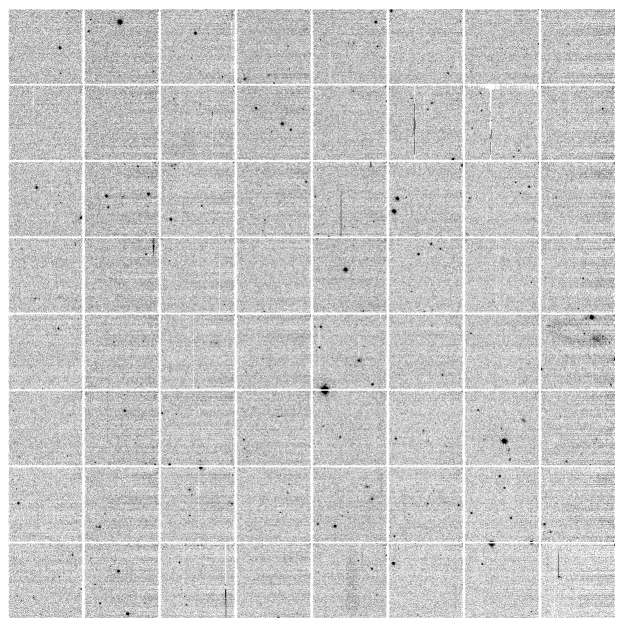}
  \end{minipage}
  \caption{{\bf Dark Correction:} An example of the dark model application to exposure o5677g0123o, OTA23 (2011-04-26, 43s \gps{} filter).  The left panel shows the image data mosaicked to the OTA level, and has had the static mask applied, the overscan subtracted, and the detector non-linearity corrected.  The right panel, shows the same exposure with the dark applied in addition to the processing shown on the left, removing the amplifier glows in the cell corners.}
  \label{fig:dark image}
\end{figure}

The dark current in the GPC1 detectors has significant variations
across each cell.  The model we make to remove this signal considers
each pixel individually, independent of any neighbors.  To construct
this model, we fit a multi-dimensional model to the array of input
pixels from a randomly selected set of 100-150 overscan and
non-linearity corrected dark frames chosen from a given date range.
The model fits each pixel as a function of the exposure time $t_{exp}$
and the detector temperature $T_{chip}$ of the input images such that
$\mathrm{dark} = a_0 + a_1 t_{exp} + a_2 T_{chip} t_{exp} + a_3
T_{chip}^2 t_{exp}$.  This fitting uses two iterations to produce a
clipped fit, rejecting at the $3\sigma$ level.  The final coefficients
$a_i$ for the dark model are stored in the detrend image.  The
constant $a_0$ term includes the residual bias signal after overscan
subtraction, and as such, a separate bias subtraction is not
necessary.

Applying the dark model is simply a matter of calculating the response
for the exposure time and detector temperature of the image to be
corrected, and subtracting the resulting dark signal from the image.
Figure \ref{fig:dark image} shows the results of the dark subtraction.

\subsubsection{Time evolution}

\begin{figure}
  \centering
  \includegraphics[width=0.9\hsize,angle=0,clip]{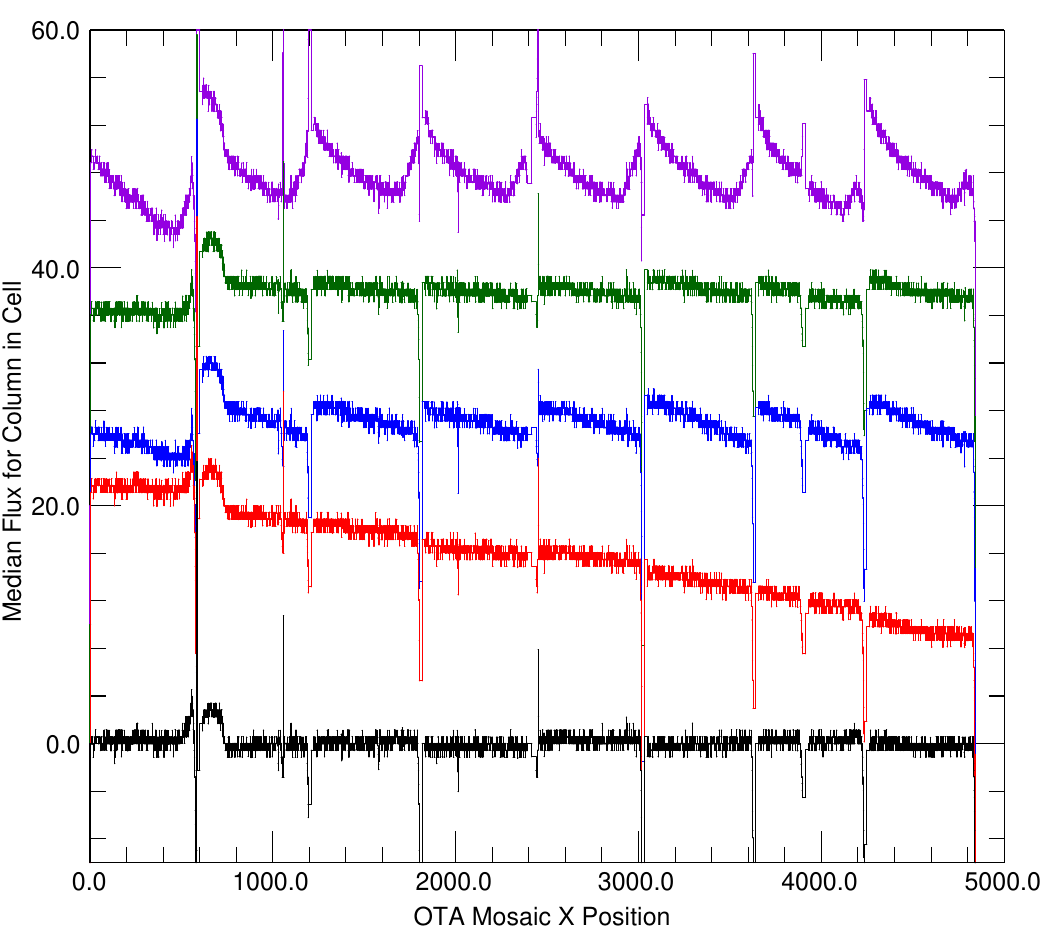}
  \caption{Example showing a profile cut across exposure o5676g0195,
    OTA67 (2011-04-25, 43s \gps{} filter).  The entire first row of
    cells (xy00-xy07) have had a median calculated along each pixel
    column on the OTA mosaicked image.  Arbitrary offsets have been
    applied so the curves do not overlap.  The top curve (in purple)
    shows the initial raw profile, with no dark model applied.  The
    next curve (in green) shows the smoother profile after applying
    the appropriate B-mode dark model.  Applying the (incorrect)
    A-mode dark instead results in the third (blue) curve, which shows
    a significant increase in gradients across the cells.  The fourth
    (red) curve is the result of applying the PATTERN.CONTINUITY
    correction along with the B-mode dark model.  Although this
    creates a larger gradient across the mosaicked images, it
    decreases the cell-to-cell boundary offsets.  The bottom (black)
    curve shows the final image profile after all detrending and
    background subtraction (no offset applied).  The bright source at
    the cell xy00 to xy01 transition is a result of a large optical
    ghost which, due to the area covered, increases the median level
    more than the field stars.}
  \label{fig:dark switching}
\end{figure}

The dark model is not consistently stable over the full survey, with
significant drift over the course of multiple months.  Some of the
changes in the dark can be attributed to changes in the voltage
settings of the GPC1 controller electronics, but the causes of others
are unknown.  We can separate the dark model history of GPC1 into
three epochs.  The first epoch covers all data taken prior to
2010-01-23.  This epoch used a different header keyword for the
detector temperature, making data from this epoch incompatible with
later dark models.  In addition, the temperatures recorded in this
value were not fully calibrated, making the dark model generated less
reliable.

The second epoch covers data between 2010-01-23 and 2011-05-01, and is
characterized by a largely stable but oscillatory dark solution.
The dark model switches between two modes apparently at
random.  No clear cause has been established for the switching, but
there are clear differences between the two modes that require the
observation dates to be split to use the model that is most
appropriate.

The initial evidence of these two modes comes from the discovery of a
slight gradient along the rows of certain cells.  This is a result of
a drift in the bias level of the detector as it is read out.  An
appropriate dark model should remove this gradient entirely.  For
these two modes, the direction of this bias drift is different, so a
single dark model generated from all dark images in the time range
over corrects the positive-gradient mode, and under corrects the
negative-gradient mode.  Upon identifying this two-mode behavior, and
determining the dates each mode was dominant, two separate dark
models were constructed from appropriate ``A'' and ``B'' mode dark
frames.  Using the appropriate dark minimizes the effect of this bias
gradient in the dark corrected data.  

The bias drift gradients of the mode switching can be visualized in
Figure \ref{fig:dark switching}.  This figure shows the image profile
along the x-pixel axis binned along the full y-axis of the first row
of cells.  The raw data is shown, illustrating the positional
dependence the dark signal has on the image values.  In addition,
both the correct B-mode dark and incorrect A-mode dark have been
applied to this image, showing that although both correct the bulk of
the dark signal, using the incorrect mode creates larger intensity
gradients.

After 2011-05-01, the two-mode behavior of the dark disappears, and is
replaced with a slow observation-date-dependent drift in the magnitude
of the gradient.  This drift is sufficiently slow that we have modeled
it by generating models for different date ranges.  These darks cover
the range from 2011-05-01 to 2011-08-01, 2011-08-01 to 2011-11-01, and
2011-11-01 and on.  The reason for this time evolution is unknown, but
as it is correctable with a small number of dark models, this does not
significantly impact detrending.

\subsubsection{Video Dark}
\label{sec:video_darks}

Individual cells on GPC1 can be repeatedly read to create a video
signal used for telescope guiding.  However, when a cell is used for
this purpose, the dark signal for the entire OTA is changed.  The most
noticeable feature of this change is that the glows in cell corners
caused by the read-out amplifiers are suppressed.  As a result, using
the standard dark model on the data for these OTAs results in
oversubtraction of the corner glow.

To generate a correction for this change, a set of video dark models
were created by running the standard dark construction process on a
series of dark frames that had the video signal enabled for some
cells.  GPC1 can only run video signals on a subset of the OTAs at a
given time.  This requires two passes to enable the video signal
across the full set of OTAs that support video cells.  This is
convenient for the process of creating darks, as those OTAs that do
not have video signals enabled create standard dark models, while the
video dark is created for those that do.

\begin{figure}
  \centering
  \begin{minipage}{0.45\hsize}
    \includegraphics[width=0.9\hsize,angle=0,clip]{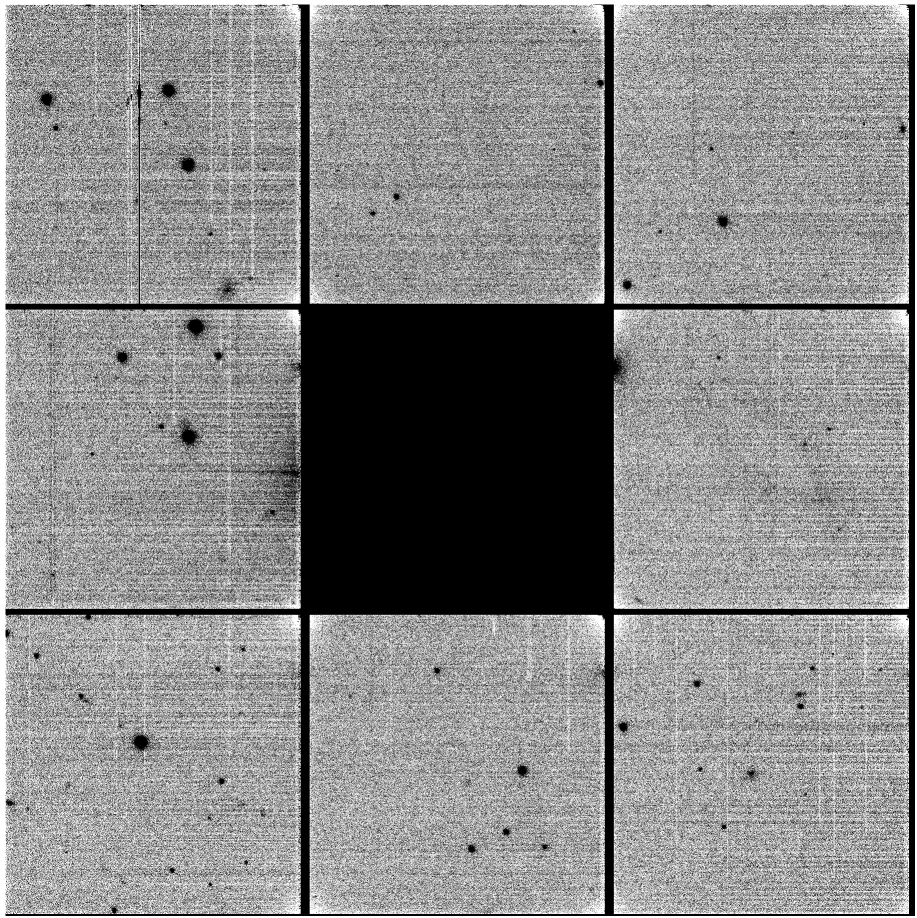}
  \end{minipage}%
  \begin{minipage}{0.45\hsize}
    \includegraphics[width=0.9\hsize,angle=0,clip]{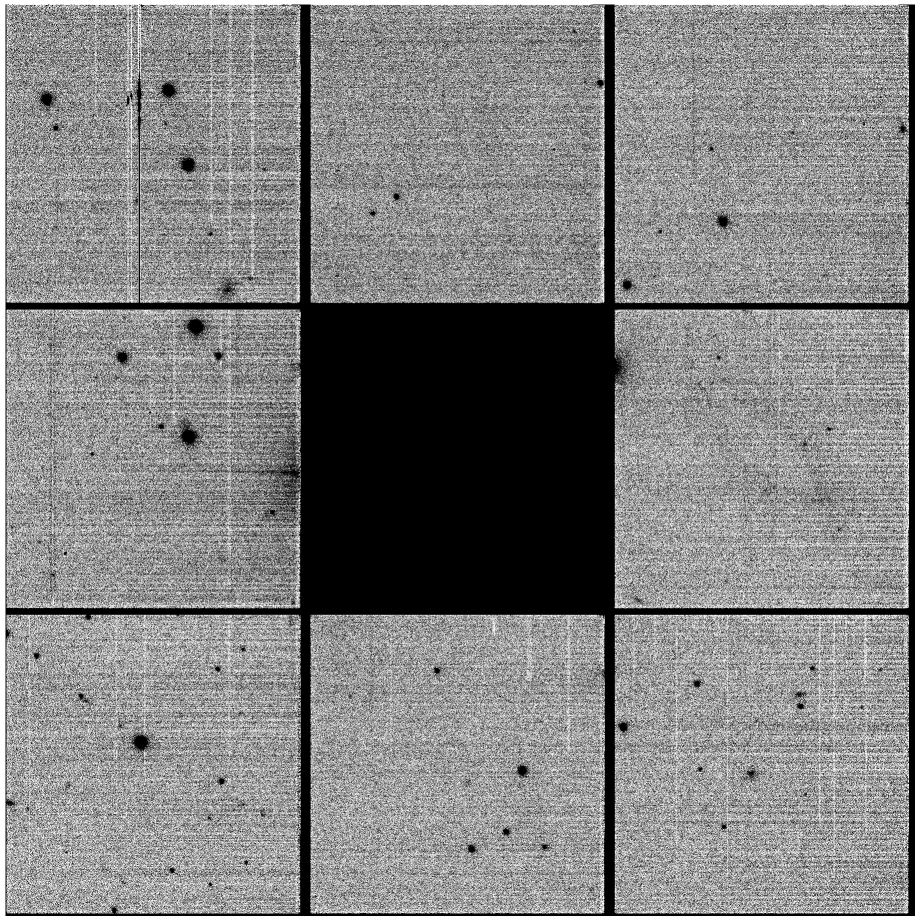}
  \end{minipage}
  \caption{{\bf Video Dark:} An example of the video dark model application to exposure o5677g0123o, OTA22 (2011-04-26, 43s \gps{} filter), which has a video cell located in cell xy16.  The left panel shows the image data mosaicked to the OTA level, and has had the static mask applied, the overscan subtracted, the detector non-linearity corrected, and a regular dark applied.  The right panel, shows the same exposure with a video dark applied instead of the standard dark.  The main impact of this change is the improved correction of the corner glows, which are over subtracted with the standard dark.}
  \label{fig:video_darks}
\end{figure}

This simultaneous construction of video and standard dark models is
useful, as it provides a way to isolate the response on the standard
dark from the video signals.  If the standard and video dark signals
are separable, then archival video darks can be constructed by
applying the video dark response to the previously constructed dark
models.  Raw video dark frame data only exists after 2012-05-16, when
this problem was initially identified, so any data prior to that can
not be directly corrected for the video dark signal.  Testing the
separability shows that constructing a video dark for older data
simply as $VD_{Old} = D_{Old} - D_{Modern} + VD_{Modern}$ produces a
satisfactory result that does not over subtract the amplifier glow.
This is shown in figure \ref{fig:video_darks}, which shows video cells
from before 2012-05-16, corrected with both the standard and video
darks, with the early video dark constructed in such a manner.

\subsection{Noisemap}
\label{sec:noisemap}

Based on a study of the positional dependence of all detected sources,
we discovered that the cells in GPC1 do not have uniform noise
characteristics.  Instead, there is a gradient along the pixel rows,
with the noise generally higher away from the read out amplifier
(higher cell $x$ pixel positions).  This is likely an effect of the
row-by-row bias issue discussed below (Section~\ref{sec:pattern.row}).
As a result of this increased noise, more sources are detected in the
higher noise regions when the read noise is assumed constant across
the readout.

To mitigate this noise gradient, we constructed an initial set of
noisemap images by measuring the median variance on bias frames
processed as science images.  The variance is calculated in boxes of
20x20 pixels, and then linearly interpolated to cover the full image.

Unfortunately, due to correlations within this noise, the variance
measured from the bias images does not fully remove the positional
dependence of objects that are detected.  This simple noisemap
underestimates the noise observed when the image is filtered during
the object detection process.  This filtering convolves the background
noise with a PSF, which has the effect of amplifying the correlated
peaks in the noise.  This amplification can therefore boost background
fluctuations above the threshold used to select real objects,
contaminating the final object catalogs.

In the detection process, we expect false positives at a low rate,
given that all sources are required to be significant at the $5\sigma$
level.  Since the observed false positive rate was significantly
higher than expected, we implemented an empirical ``boost'' to
increase the noisemap to more accurately account for the position
dependent read noise.  By binning the number of false positives
measured on the bias frames on the noisemap inputs using 20 pixel
boxes in the cell $x$-axis, and comparing this to the number expected
from random Gaussian noise, we estimated the true read noise level.

As the noisemap uses bias frames that have had a dark model
subtracted, we constructed noisemaps for each dark model used for
science processing.  There is some evidence that the noise has changed
over time as measured on full cells, so matching the noisemap to the
dark model allows for these changes to be tracked.  There is no
evidence that the noisemap has the A/B modes found in the dark, so we
do not generate separate models for that time period.

The noisemap detrend is not directly applied to the science image.
Instead, it is used to construct the weight image that contains the
pixel-by-pixel variance for the \IPPstage{chip} stage image.  The
initial weight image is constructed by dividing the science image by
the cell gain (approximately 1.0 e$^{-} /$ DN).  This weight image
contains the expected Poissonian variance in electrons measured.  The
square of the noisemap is then added to this initial weight, adding
the additional empirical variance term in place of a single read noise
value.

\subsection{Flat}

Determining a flat field correction for GPC1 is a challenging
endeavor, as the wide field of view makes it difficult to construct a
uniformly illuminated image.  Using a dome screen is not possible, as
the variations in illumination and screen rigidity create large
scatter between different images that are not caused by the detector
response function.  Because of this, we use sky flat images taken at
twilight, which are more consistently illuminated than screen flats.
We calculate the mean of these images to determine the initial flat
model.

From this starting skyflat model, we construct a photometric
correction to remove the effect of the illumination differences over
the detector surface.  This is done by dithering a series of science
exposures with a given pointing, as described in
\citet{2004PASP..116..449M}.  By fully calibrating these exposures
with the initial flat model, and then comparing the measured fluxes
for the same star as a function of position on the detector, we can
determine position dependent scaling factors.  From the set of scaling
factors for the full catalog of stars observed in the dithered
sequence, we can construct a model of the error in the initial flat
model as a function of detector position.  Applying a correction that
reduces the amplitude of these errors produces a flat field model that
better represents the true detector response.

In addition to this flat field applied to the individual images, the
``ubercal'' analysis -- in which photometric data are used define
image zero points
\citep[][]{2012ApJ...756..158S,magnier2017.calibration} and in turn
used used to calibrate the database of all detections -- constructs
``in catalog'' flat field corrections.  Although a single set of image
flat fields was used for the PV3 processing of the entire $3\pi$
survey, five separate ``seasons'' of database flat fields were needed
to ensure proper calibration.  This indicates that the flat field
response is not completely fixed in time.  More details on this
process are contained in \citet{magnier2017.calibration}.

\subsection{Fringe correction}
\label{sec:fringe}
% det_id 296 is the fringe we use.

Due to variations in the thickness of the detectors, we observe
interference patterns at the infrared end of the filter set, as the
wavelength of the light becomes comparable to the thickness of the
detectors.  Visually inspecting the images shows that the fringing is
most prevalent in the \yps{} filter images, with negligible fringing in the
other bands.  As a result of this, we only apply a fringe correction
to the \yps{} filter data.

The fringe used for PV3 processing was constructed from a set of 20
120s science exposures.  These exposures are overscan subtracted, and
corrected for non-linearity, and have the dark and flat models
applied.  These images are smoothed with a Gaussian kernel with
$\sigma = 2$ pixels to minimize pixel to pixel noise.  The fringe
image data is then constructed by calculating the clipped mean of the
input images with two iteration of clipping at the $3\sigma$ level.

A coarse background model for each cell is constructed by calculating
the median on a 3x3 grid (approximately 200x200 pixels each).  A set
of 1000 points are randomly selected from the fringe image for each
cell, and a median calculated for this position in a 10x10 pixel box,
with the background level subtracted.  These sample locations provide
scale points to allow the amplitude of the measured fringe to be
compared to that found on science images.

\begin{figure}
  \centering
  \begin{minipage}{0.45\hsize}
    \includegraphics[width=0.9\hsize,angle=0,clip]{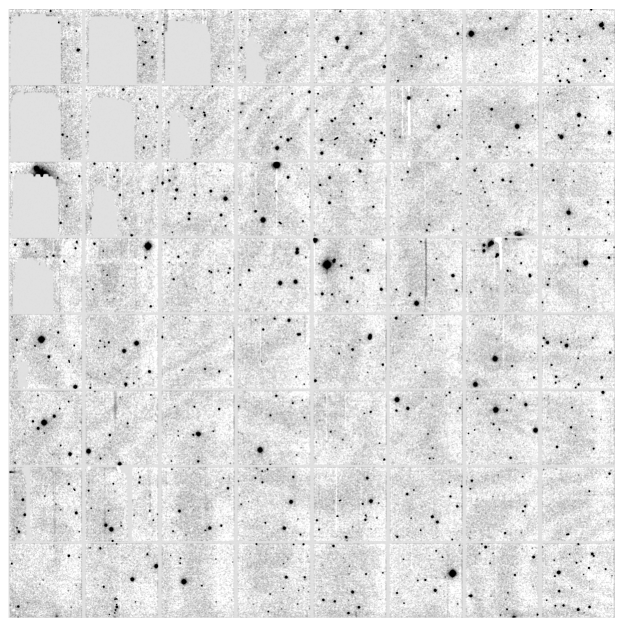}
  \end{minipage}%
  \begin{minipage}{0.45\hsize}
    \includegraphics[width=0.9\hsize,angle=0,clip]{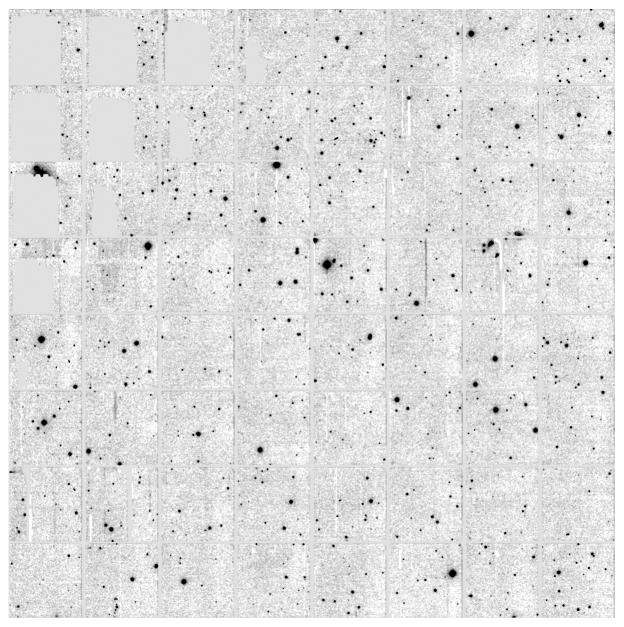}
  \end{minipage}
  \caption{{\bf Fringing:} Example of the \yps{} filter fringe pattern
    on exposure o5220g0025o OTA53 (\yps{} filter 30s).  The left panel
    shows the OTA mosaic with all detrending except the fringe
    correction, while the right shows the same including the fringe
    correction.  Both images have been smoothed with a Gaussian with
    $\sigma = 3$ pixels to highlight the faint and large scale fringe
    patterns.  }
  \label{fig: fringe example}
\end{figure}

To apply the fringe, the same sample locations are measured on the
science image to determine the relative strength of the fringing in
that particular image.  A least squares fit between the fringe
measurements and the corresponding measurements on the science image
provides the scale factor multiplied to the fringe before it is
subtracted from the science image.  An example of the fringe correction can be seen in Figure~\ref{fig: fringe example}.  

\subsection{Masking}
\label{sec:masking}

\subsubsection{Static Masks}
\label{sec:static_masks}

Due to the large size of the detector, it is expected that there are a
number of pixels that respond poorly.  To remove these pixels, we have
constructed a mask that identifies the known defects.  This mask is
referred to as the ``static'' mask, as it is applied to all images
processed.  The ``dynamic'' mask (Section \ref{sec:dynamic_masks}) is
calculated based on objects in the field, and so changes between
images.  Construction of the static mask consists of three phases.

First, regions in which the charge transfer efficiency (CTE) is low
compared to the rest of the detector are identified.  Twenty-five of
the sixty OTAs in GPC1 show some evidence of poor CTE, with this
pattern appearing (to varying degrees) in roughly triangular patches.
During the manufacture of the devices, an improperly tuned
semiconductor process step resulted in a radial pattern of poor
performance on some silicon wafers.  When the OTAs were cut from these
wafers, the outer corners exhibited the issue.  To generate the mask
for these regions, a sample set of 26 evenly-illuminated flat-field
images were measured to produce a map of the image variance in 20x20
pixel bins.  As the flat screen is expected to illuminate the image
uniformly on this scale, the expected variances in each bin should be
Poissonian distributed with the flux level.  However, in regions with
poor CTE, adjacent pixels are not independent, as the charge in those
pixels is more free to spread along the image columns.  This reduces
the pixel-to-pixel differences, resulting in a lower than expected
variance.  All regions with variance less than half the average image
level are added to the static mask.

The next step of mask construction is to examine the flat and dark
models, and exclude pixels that appear to be poorly corrected by these
models.  The DARKMASK process looks for pixels that are more than
$8\sigma$ discrepant in $10\%$ of the 100 input dark frame images
after those images have had the dark model applied to them.  These
pixels are assumed to be unstable with respect to the dark model, and
have the DARK bit set in the static mask, indicating that they are
unreliable in scientific observing.  Similarly, the FLATMASK process
looks for pixels that are $3\sigma$ discrepant in the same fraction of
16 input flat field images after both the dark and flat models have
been applied.  Those pixels that do not follow the flat field model of
the rest of image are assigned the FLAT mask bit in the static mask,
removing the pixels that cannot be corrected to a linear response.

% http://svn.pan-starrs.ifa.hawaii.edu/trac/ipp/wiki/StaticMasks20101215
The final step of mask construction is to examine the detector for
bright columns and other static pixel issues.  This is first done by
processing a set of 100 \ips{} filter science images in the same fashion as
for the DARKMASK.  A median image is constructed from these inputs
along with the per-pixel variance.  These images are used to identify
pixels that have unexpectedly low variation between all inputs, as
well as those that significantly deviate from the global median value.
Once this initial set of bad pixels is identified, a $3\times{}3$
pixel triangular kernel is convolved with the initial set, and any
convolved pixel with value greater than 1 is assigned to the static
mask.  This does an excellent job of removing the majority of the
problem pixels.  A subsequent manual inspection allows human
interaction to identify other inconsistent pixels including the
vignetted regions around the edge of the detector.  

\begin{figure}
  \centering
  \includegraphics[width=0.9\hsize,angle=0,clip]{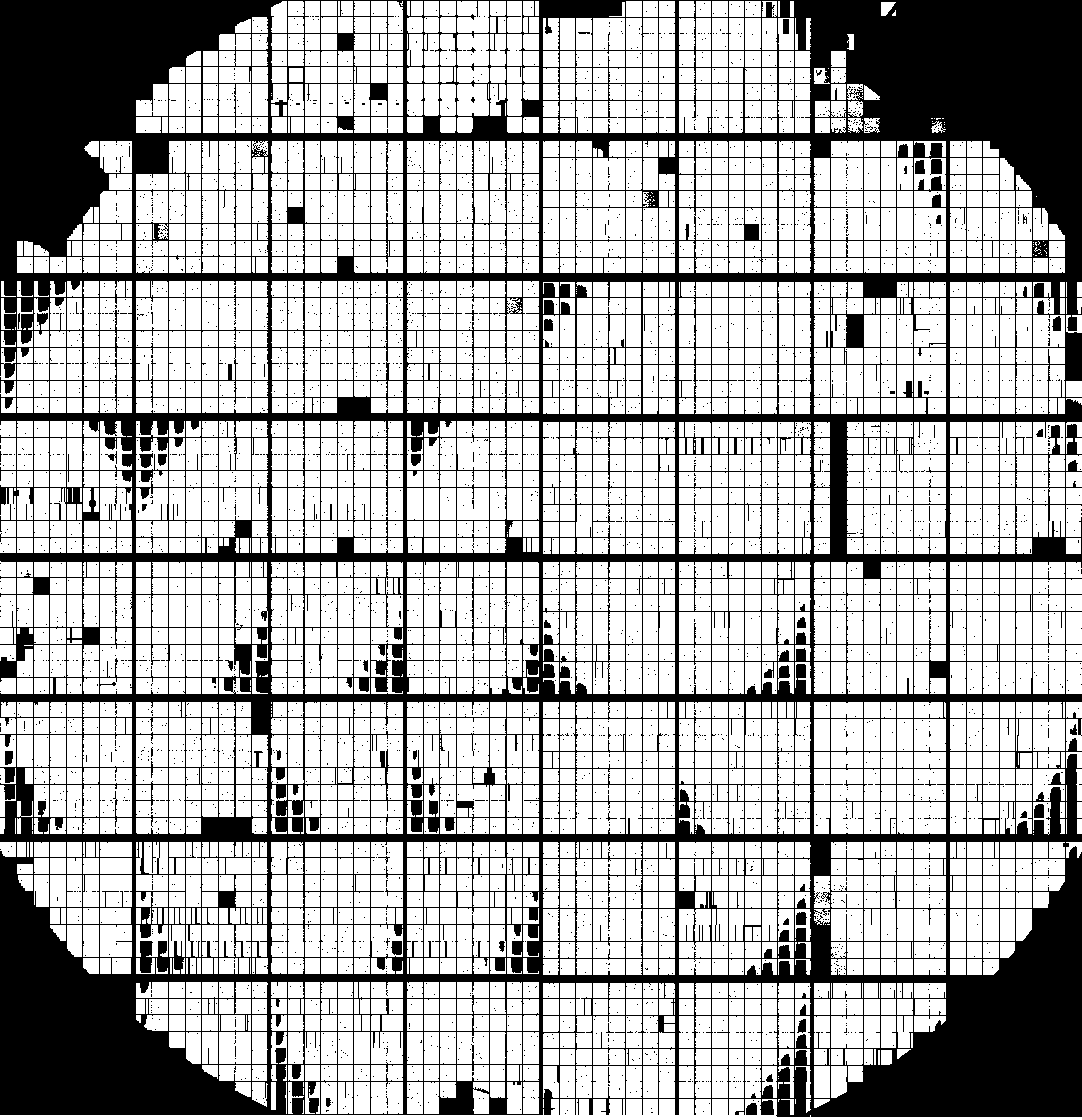}
  \caption{Image map of the GPC1 static mask.  The CTE regions are clearly visible as roughly triangular patches covering the corners of some OTAs.  Some entire cells are masked, including an entire column of cells on OTA14.  Calcite cells remove large areas from OTA17 AND OTA76.}
  \label{fig:static mask}
\end{figure}

Figure \ref{fig:static mask} shows an example of the static mask for
the full GPC1 field of view.  Table \ref{tab:mask_values} lists the
bit mask values used for the different sources of masking.

\begin{deluxetable*}{ccl}
  \tablecolumns{3}
  \tablewidth{0pc}
  \tablecaption{GPC1 Mask Values}
  \tablehead{\colhead{Mask Name} & \colhead{Mask Value} &
    \colhead{Description (static values listed in bold)}}
  \startdata
  {\bf DETECTOR } & {\bf 0x0001}  & {\bf A detector defect is present.} \\
  {\bf FLAT     } & {\bf 0x0002}  & {\bf The flat field model does not calibrate the pixel reliably.} \\
  {\bf DARK     } & {\bf 0x0004}  & {\bf The dark model does not calibrate the pixel reliably.} \\
  {\bf BLANK    } & {\bf 0x0008}  & {\bf The pixel does not contain valid data.} \\
  {\bf CTE      } & {\bf 0x0010}  & {\bf The pixel has poor charge transfer efficiency.} \\
  SAT      & 0x0020 & The pixel is saturated. \\
  LOW      & 0x0040 & The pixel has a lower value than expected. \\
  SUSPECT  & 0x0080 & The pixel is suspected of being bad (overloaded with the BURNTOOL bit). \\
  BURNTOOL & 0x0080 & The pixel contain an burntool repaired streak. \\
  CR       & 0x0100 & A cosmic ray is present. \\
  SPIKE    & 0x0200 & A diffraction spike is present. \\
  GHOST    & 0x0400 & An optical ghost is present. \\
  STREAK   & 0x0800 & A streak is present. \\
  STARCORE & 0x1000 & A bright star core is present. \\
  CONV.BAD & 0x2000 & The pixel is bad after convolution with a bad pixel. \\
  CONV.POOR& 0x4000 & The pixel is poor after convolution with a bad pixel. \\
  MARK     & 0x8000 & An internal flag for temporarily marking a pixel. \\
  \enddata
  \label{tab:mask_values}
\end{deluxetable*}

\subsubsection{Dynamic masks}
\label{sec:dynamic_masks}

In addition to the static mask that removes the constant detector
defects, we also generate a set of dynamic masks that change with the
astronomical features in the image.  These masks are advisory in
nature, and do not completely exclude the pixel from further
processing consideration.  The first of these dynamic masks is the
burntool advisory mask described below.  These pixels are included for
photometry, but are rejected more readily in the stacking and
difference image construction, as they are more likely to have small
deviations due to imperfections in the burntool correction.

The remaining dynamic masks are generated in the IPP \IPPstage{camera}
stage, at which point all object photometry is complete, and an
astrometric solution is known for the exposure.  This added
information provides the positions of bright sources based on the
reference catalog, including those that fall slightly out of the
detector field of view or within the inter chip gaps, where internal
photometry may not identify them.  These bright sources are the origin
for many of the image artifacts that the dynamic mask identifies and
excludes.

\paragraph{Electronic crosstalk ghosts}
\label{sec:crosstalk}

Due to electrical crosstalk between the flex cables connecting the
individual detector OTA devices, ghost objects can be created by the
presence of a bright source at a different position on the camera.
Table \ref{tab:crosstalk_rules} summarizes the list of known crosstalk
rules, with an estimate of the magnitude difference between the source
and ghost.  For all of the rules, any source cell $v$ within the
specified column of cells on any of the OTAs in the specified column
of OTAs $Y$ can create a ghost in the same cell $v$ and OTA $Y$ in the
target column of cells and OTAs.  This effect depends on the number of
electrons detected for the star, thus the size of the ghost scales
with the instrumental magnitude ($m_{inst} = -2.5 \log_{10} (ADU)$) of
the star.  In each of these cases, a source object with $m_{inst} <
-14.47$) (corresponding to $\rps \lesssim 14$ for the $3\pi$ survey)
creates a ghost object many orders of magnitude fainter at the target
location.  The cell ($x,y$) pixel coordinate is identical between
source and ghost, as a result of the transfer occurring as the devices
are read.  A circular mask is added to the ghost location with radius
$R = 3.44 \left(-14.47 - m_{inst,source}\right)$ pixels; only
positive radii are allowed.  Any objects in the photometric catalog
found at the location of the ghost mask have the GHOST mask bit set,
marking the object as a likely ghost.  The majority of the crosstalk
rules are bi-directional, with a source in either position creating a
ghost at the corresponding crosstalk target position.  The two
faintest rules are uni-directional, due to differences in the
electronic path for the crosstalk.

For the very brightest sources ($m_{inst} < -15$), there can be
crosstalk ghosts between all columns of cells during the readout.
These ``bleed'' ghosts were originally identified as ghosts of the
saturation bleeds appearing in the neighboring cells, and as such, the
masking for these objects puts a rectangular mask down from top to
bottom of cells in all columns that are in the same row of cells as
the bright source.  The width of this box is a function of the source
magnitude, with $W = 5 \times \left(-15 - m_{inst,source}\right)$
pixels.

\paragraph{Optical ghosts}
\label{sec:optical_ghosts}

\begin{figure}
  \centering
  \includegraphics[width=0.9\hsize,angle=0,clip]{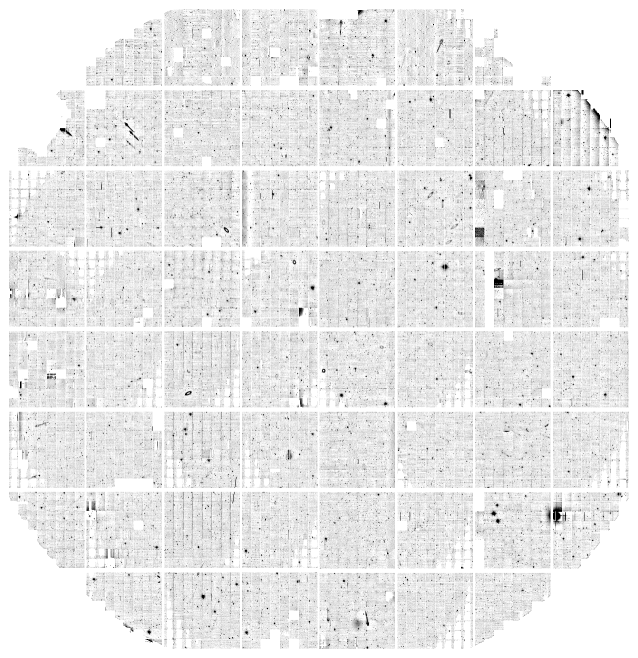}
  \caption{{\bf Ghosts:} Example of the full GPC1 field of view
    illustrating the sources and destinations of optical ghosts on
    exposure o5677g0123o (2011-04-26, 43s \gps{} filter).  The bright
    stars on OTA33 and OTA44 result in nearly circular ghosts on the
    opposite OTA.  In contrast, the trio of stars on OTA11 result in
    very elongated ghosts on OTA66.}
  \label{fig:optical ghosts}
\end{figure}

The anti-reflective coating on the optical surfaces of GPC1 is less
effective at shorter wavelengths, which can allow bright sources to
reflect back onto the focal plane and generate large out-of-focus
objects.  Due to the wavelength dependence, these objects are most
prominent in the \gps{} filter data.  These objects are the result of
light reflecting back off the surface of the detector, reflecting
again off the lower surfaces of the optics (particularly the L1
corrector lens), and then back down onto the focal plane.  Due to the
extra travel distance, the resulting source is out of focus and
elongated along the radial direction of the camera focal
plane. Figure~\ref{fig:optical ghosts} shows an example exposure with
several prominent optical ghosts.

These optical ghosts can be modeled in the focal plane coordinates
($L,M$) which has its origin at the center of the focal plane.  In
this system, a bright object at location ($L,M$) on the focal plane
creates a reflection ghost on the opposite side of the optical axis
near ($-L,-M$).  The exact location is fit as a third order polynomial
in the focal plane $L$ and $M$ directions (as listed in Table
\ref{tab:ghost_centers}).  An elliptical annulus mask is constructed
at the expected ghost location, with the major and minor axes of the inner and outer elliptical annuli defined
by linear functions of the ghost distance from the optical axis, and
oriented with the ellipse major axis is along the radial direction
(Table \ref{tab:ghost_radii}).  All stars brighter than a
filter-dependent threshold (listed in Table
\ref{tab:ghost_magnitudes}) have such masks constructed.

\begin{deluxetable}{lllc}
  \tablecolumns{4}
  \tablewidth{0pc}
  \tablecaption{GPC1 Crosstalk Rules}
  \tablehead{\colhead{Type}&\colhead{Source OTA/Cell}&\colhead{Ghost OTA/Cell}&\colhead{$\Delta m$}}
  \startdata
  Inter-OTA & OTA2Y XY3v & OTA3Y XY3v & 6.16 \\
            & OTA3Y XY3v & OTA2Y XY3v &      \\
            & OTA4Y XY3v & OTA5Y XY3v &      \\
            & OTA5Y XY3v & OTA4Y XY3v &      \\
  Intra-OTA & OTA2Y XY5v & OTA2Y XY6v & 7.07 \\
            & OTA2Y XY6v & OTA2Y XY5v &      \\
            & OTA5Y XY5v & OTA5Y XY6v &      \\
            & OTA5Y XY6v & OTA5Y XY5v &      \\
  One-way   & OTA2Y XY7v & OTA3Y XY2v & 7.34 \\
            & OTA5Y XY7v & OTA4Y XY2v &      \\
  \enddata
  \label{tab:crosstalk_rules}
\end{deluxetable}

\begin{deluxetable}{lcc}
  \tablecolumns{3}
  \tablewidth{0pc}
  \tablecaption{Optical Ghost Center Transformations}
  \tablehead{\colhead{Polynomial Term}&\colhead{$L$ center}&\colhead{$M$ center}}
  \startdata 
  $x^0 y^0$ & -1.215661e+02 &  2.422174e+01 \\
  $x^1 y^0$ &  1.321875e-02 &  4.170486e-04 \\
  $x^2 y^0$ & -4.017026e-09 & -1.934260e-08 \\
  $x^3 y^0$ &  1.148288e-10 & -1.173657e-12 \\
  $x^0 y^1$ & -1.908074e-03 &  1.189352e-02 \\
  $x^1 y^1$ &  8.479150e-08 & -9.256748e-08 \\
  $x^2 y^1$ &  1.635732e-11 &  1.140772e-10 \\
  $x^0 y^2$ &  2.625405e-08 &  8.123932e-08 \\
  $x^1 y^2$ &  1.125586e-10 &  1.328378e-11 \\
  $x^0 y^3$ &  2.912432e-12 &  1.170865e-10 \\
  \enddata
  \label{tab:ghost_centers}
\end{deluxetable}

\begin{deluxetable*}{lcccc}
  \tablecolumns{5}
  \tablewidth{0pc}
  \tablecaption{Optical Ghost Annulus Axis Length}
  \tablehead{\colhead{Radial Order}&\colhead{Inner Major Axis}&\colhead{Inner Minor Axis}&\colhead{Outer Major Axis}&\colhead{Outer Minor Axis}}
  \startdata
  $r^0$ & 3.926693e+01 & 5.287548e+01 & 7.928722e+01 & 1.314265e+02 \\
  $r^1$ & 5.325759e-03 &-2.191669e-03 & 1.722181e-02 & -2.627153e-03 \\
  \enddata
  \label{tab:ghost_radii}
\end{deluxetable*}

%% \begin{deluxetable}{lcccc}
%%   \tablecolumns{5}
%%   \tablewidth{0pc}
%%   \tablecaption{Optical Ghost Annulus Axis Length}
%%   \tablehead{\colhead{Order}&\colhead{Maj$_{\rm in}$}&\colhead{Min$_{\rm in}$}&    \colhead{Maj$_{\rm out}$}&\colhead{Min$_{\rm out}$}}
%%   \startdata
%%   $r^0$ & 3.926693e+01 & 5.287548e+01 & 7.928722e+01 & 1.314265e+02 \\
%%   $r^1$ & 5.325759e-03 &-2.191669e-03 & 1.722181e-02 & -2.627153e-03 \\
%%   \enddata
%%   \label{tab:ghost_radii}
%% \end{deluxetable}

\begin{deluxetable}{lrr}
  \tablecolumns{3}
  \tablewidth{0pc}
  \tablecaption{Optical Ghost Magnitude Limits}
% \tablehead{\colhead{Filter} & \colhead{$m_{inst}$} & \colhead{\parbox{2cm}{Apparent mag ($3\pi$)}}}
  \tablehead{\colhead{Filter} & \colhead{$m_{inst}$} & \colhead{Apparent mag ($3\pi$)}}
  \startdata
  \gps{} & -16.5 & 12.2 \\
  \rps{} & -20.0 &  8.9 \\
  \ips{} & -25.0 &  3.7 \\
  \zps{} & -25.0 &  3.4 \\
  \yps{} & -25.0 &  2.5 \\
  \wps{} & -20.0 & 10.2 \\
  \enddata
  \label{tab:ghost_magnitudes}
\end{deluxetable}

\paragraph{Optical glints}
\label{sec:glints}

\begin{figure}
  \centering
  \includegraphics[width=0.9\hsize,angle=0,clip]{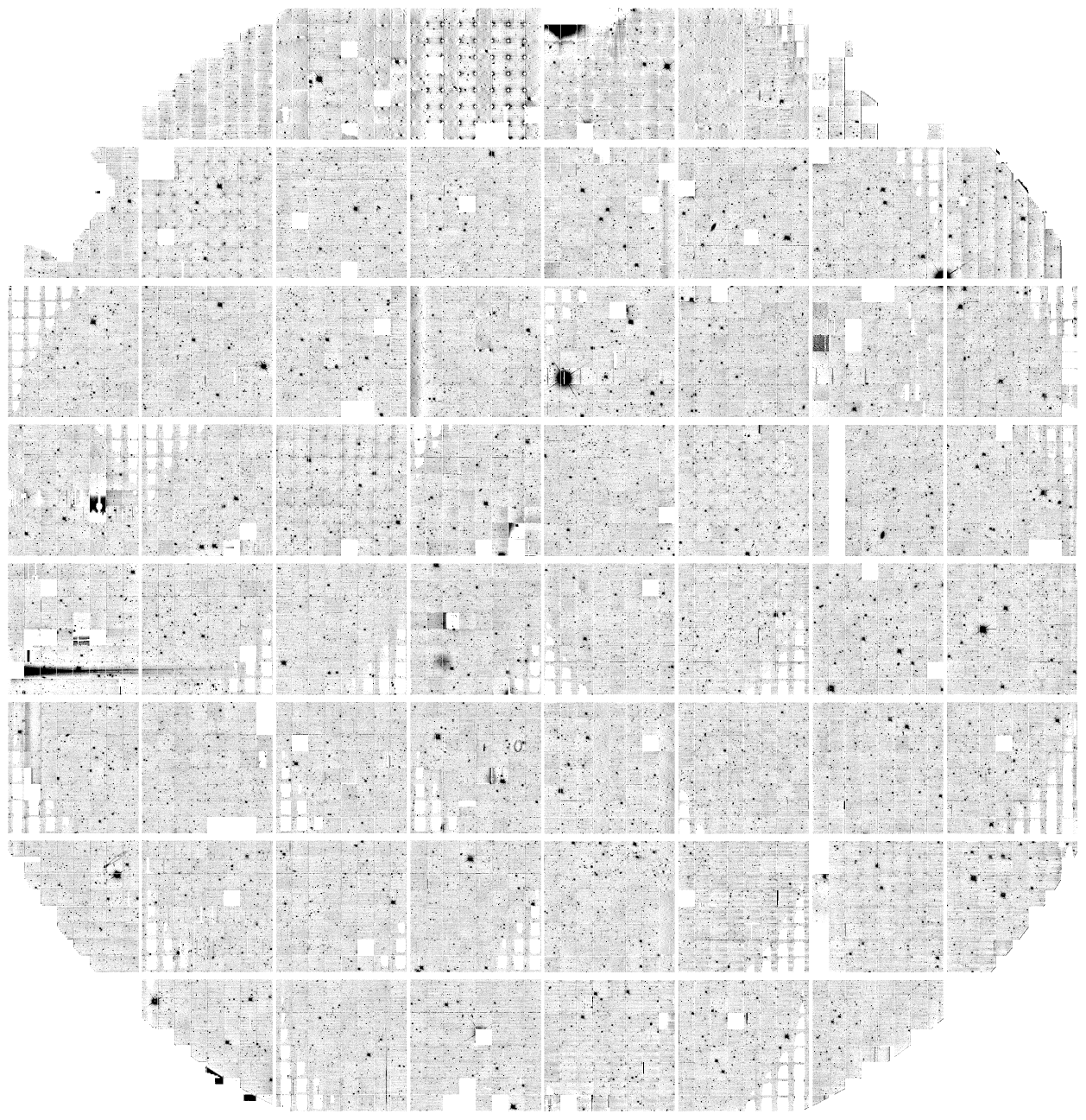}
  \caption{{\bf Glints:}  Example of a glint on exposure o5379g0103o (2010-07-02, 45s \ips{} filter).  The source star out of the field of view creates a long reflection that extends through OTA73 and OTA63.}
  \label{fig:optical glints}
\end{figure}

Prior to 2010-08-24, a reflective surface at the edge of the camera
aperture was incompletely screened to light passing through the
telescope.  Sources brighter than $m_{inst} = -21$ ($\rps \lesssim
7.5$) that fell on this reflective surface resulted in light being
scattered across the detector surface in a long narrow glint.  
Figure~\ref{fig:optical glints} shows an example exposure with
a prominent optical glint.

This reflective surface in the camera was physically masked on
2010-08-24, removing the possibility of glints in subsequent data, but
images that were taken prior to this date have an advisory dynamic
mask constructed when a reference source falls on the focal plane
within one degree of the detector edge.  This mask is 150 pixels wide,
with length $L = 2500 \left(-20 - m_{inst}\right)$ pixels.  These
glint masks are constructed by selecting sufficiently bright sources
in the reference catalog that fall within rectangular regions around
each edge of the GPC1 camera.  These regions are separated from the
edge of the camera by 17 arcminutes, and extend outwards an additional
degree.

\paragraph{Diffraction Spikes and Saturated Stars}
\label{sec:diffraction_spikes}

Bright sources also form diffraction spikes that are dynamically
masked.  These are filter independent, and are modeled as rectangles
with length $L = 10^{0.096 \times (7.35 - m_{inst})} - 200$ and
width $W = 8 + (L - 200) \times 0.01$, with negative values indicating no
mask is constructed, as the source is likely too faint to produce the
feature.  These spikes are dependent on the camera rotation, and are
oriented based on the header keyword at $\theta = n \times \frac{\pi}{2} -
\mathrm{ROTANGLE} + 0.798$, for $n = {0,1,2,3}$.

\begin{figure}
  \centering
  \includegraphics[width=0.9\hsize,angle=0,clip]{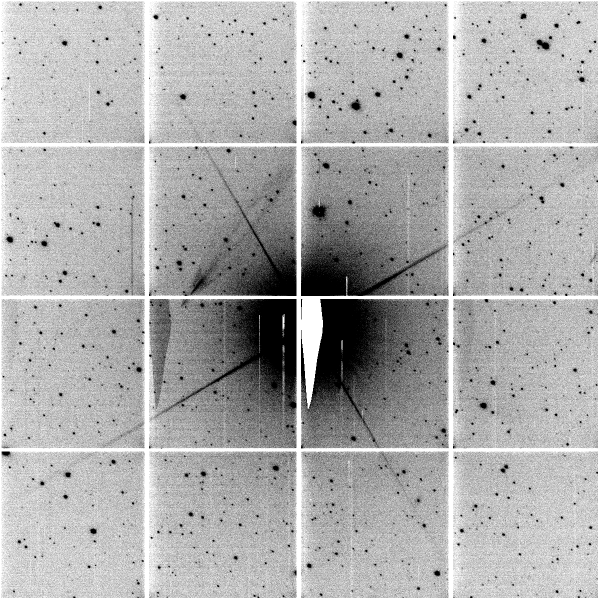}
  \caption{Example of saturated star, with diffraction spikes extending from the core on exposure o6802g0338o, OTA51 (2014-05-25, 45s \gps{} filter).}
  \label{fig:saturated star}
\end{figure}

The cores of stars that are saturated are masked as well, with a
circular mask radius $r = 10.15 \times (-15 - m_{inst})$.  An
example of a saturated star, with the masked regions for the
diffraction spikes and core saturation highlighted, is shown in Figure
\ref{fig:saturated star}.

Saturation for the GPC1 detectors varies from chip to chip and cell to
cell.  Saturation levels have been measured in the lab for each cell
and are recorded in the headers.  The IPP analysis code reads the
header value to determine the appropriate saturation point.  Of the
3840 cells in GPC1, the median saturation level is 60,400; 95\% have
saturation levels $> 54,500$ DN; 99\% have saturation levels $>
41,000$ DN.  A small number of cells have recorded saturation values
much lower than these values, but these also tend to be the cells for
which other cosmetic effects (\eg, CTE \& dark current) are strong,
likely affecting the measurement of the saturation value.

\subsubsection{Masking Fraction}
\label{sec:masking_fraction}

The GPC1 camera was designed such that where possible, OTAs with CTE
issues were placed towards the edge of the detector.  Because of this,
the main analysis of the mask fraction is based not on the total
footprint of the detector, but upon a circular reference field of view
with a radius of 1.5 degrees.  This radius corresponds approximately
to half the width and height of the detector.  This field of view
underestimates the unvignetted region of GPC1.  A second ``maximum''
field of view is also used to estimate the mask fraction within a
larger 1.628 degree radius.  This larger radius includes far larger
missing fractions due to the circular regions outside region populated
with OTAs, but does include the contribution from well-illuminated
pixels that are ignored by the reference radius.

The results of simulating the footprint of the detector as a grid of
uniformly sized pixels of $0\farcs{}258$ size are provided in Table
\ref{tab:mask fraction}.  Both fields of view contain circular
segments outside of the footprint of the detector, which increase the
area estimate that is unpopulated.  This category also accounts for
the inter-OTA and inter-cell gaps.  The regions with poor CTE also
contribute to a significant fraction of the masked pixels.  The
remaining mask category accounts for known bad columns, cells that do
not calibrate well, and vignetting.  There are also a small fraction
that have static advisory masks marked on all images.  These masks
mark regions where bright columns on one cell periodically create
cross talk ghosts on other cells.

During the \IPPstage{camera} processing, a separate estimate of the
mask fraction for a given exposure is calculated by counting the
fraction of pixels with static, dynamic, and advisory mask bits set
within the two field of view radii.  The static mask fraction is then
augmented by an estimate of the unpopulated inter-chip gaps (as the
input masks already account for the inter-cell gaps).  This estimate
does not include the circular segments outside of the detector
footprint.  This difference is minor for the reference field of view
(1\% difference), but underestimates the static mask fraction for the
maximum radius by 7.3\%.  This analysis provides the observed dynamic
and advisory mask fractions, which are 0.03\% and 3\% respectively.
The significant advisory value is a result of applying such masks to
all burntool corrected pixels.

\begin{deluxetable}{lcc}
  \tablecolumns{3}
  \tablewidth{0pc}
  \tablecaption{Mask Fraction by Mask Source}
  \tablehead{\colhead{Mask Source}&\colhead{3 Degree FOV}&\colhead{3.25 Degree FOV}}
  \startdata
  Good pixel      & 78.9\% & 71.1\% \\
  Unpopulated     & 13.1\% & 19.6\% \\
  CTE issue       &  2.3\% &  2.6\% \\
  Other issue     &  5.4\% &  6.4\% \\
  Static advisory &  0.3\% &  0.3\% \\
  \enddata
  \label{tab:mask fraction}
\end{deluxetable}

\subsection{Background subtraction}
\label{sec:background}

Once all other detrending is done, the pixels from each cell are
mosaicked into the full $4846\times{}4868$ pixel OTA image.  A
background model for the full OTA is then determined prior to the
photometric analysis.  The mosaicked image is subdivided into
$800\times{}800$ pixel segments that define each superpixel of the
background model, with the superpixels centered on the image center
and overlapping adjacent superpixels by 400 pixels.  These overlaps
help smooth the background model, as adjacent model pixels share input
pixels.

From each segment, 10000 random unmasked pixels are drawn.  In the
case where the mask fraction is large (such as on OTAs near the edge
of the field of view), and there are insufficient unmasked pixels to
meet this criterion, all possible unmasked pixels are used instead.
If this number is still small (less than 100 good pixels), the
superpixel does not have a background model calculated.  Instead,
the value assigned to that model pixel is set as the average of the
adjacent model pixels.  This allows up to eight neighboring background
values to be used to patch these bad pixels.

For the subdivisions that have sufficient unmasked pixels for the
background to be measured, the pixel values are used to calculate a
set of robust statistics for the initial background guess.  The
minimum and maximum of the values are found, and checked to ensure
that these are not the same value, which would indicate some problem
with the input values.  The values are then inserted into a histogram
with 1000 bins between the minimum and maximum values.  If the bin
with the most input pixels contains more than half of the input
values, the bin size is too coarse for the population of interest.  In
this case, a new histogram is constructed using a range corresponding
to the 20 bins closes to the peak, again dividing the range into 1000
bins.  This process is iterated up to 20 times until a binsize is
determined.  A cumulative distribution is then constructed from the
histogram, which saves the computational cost of sorting all the input
values.  The bins containing the 50-percentile point, as well as the
15.8\%, 84.1\% ($\pm 1 \sigma$), 30.8\%, 69.1\% ($\pm 0.5 \sigma$),
2.2\%, and 97.7\% ($\pm 2 \sigma$) points are identified in this
cumulative histogram.  These bins, and the two bins to either side are
then linearly interpolated to identify the pixel value corresponding
to these points in the distribution.  The 50\% point is set as the
median of the pixel distribution, with the standard deviation of the
distribution set as the median of the $\sigma$ values calculated from
the $0.5 \times (\sigma_{+1} - \sigma_{-1})$, $\sigma_{+0.5} -
\sigma_{-0.5}$, and $0.25 \times (\sigma_{+2} - \sigma_{-2})$ differences.
If this measured standard deviation is smaller than 3 times the bin
size, then all points more than 25 bins away from the calculated
median are masked, and the process is repeated with a new 1000 bin
histogram until the bin size is sufficiently small to ensure that the
distribution width is well sampled.  Once this iterative process
converges, or 20 iterations are run, the 25- and 75-percentile values
are found by interpolating the 5 bins around the expected bin as well,
and the count of the number of input values within this inner
50-percentile region, $N_{50}$, is calculated.

These initial statistics are then used as the starting guesses for a
second calculation of the background level that attempts to fit the
distribution with a Gaussian.  All pixels that were masked in the
initial calculation are unmasked, and a histogram is again constructed
from the values, with a bin size set to $\sigma_{guess} / \left( N_{50} /
500 \right)$.  With this bin size, we expect that a bin at $\pm 2
\sigma$ will have approximately 50 input points, which gives a
Poissonian signal-to-noise estimate around 7.  In the case where
$N_{50}$ is small (due to a poorly populated input image), this bin
size is fixed to be no larger than the guess of the standard
deviation.  The endpoints of the histogram are clipped based on the
input guesses, such that any input point with a value more than $5
\sigma_{guess}$ away from the input mean are excluded from
consideration.  

Two second order polynomial fits are then performed to the logarithm
of the histogram counts set at the midpoint of each bin.  The first
fit considers the lower portion of the distribution, under the
assumption that deviations from a normal distribution are caused by
real astrophysical sources that will be brighter than the true
background level.  From the bin with most pixel values, the lower
bound is set by searching for the first bin from the peak that has
fewer inputs than 25\% of the peak.  A similar search is performed for
the upper bound, but with a criterion that the bin has fewer than 50\%
of the peak.  On both sides of the peak, the bounds are adjusted to
ensure that at least seven bins, equally distributed around the peak,
are used.  The second fit is symmetric, fitting both sides of the
distribution out to the point where the bin contains fewer than 15\%
of the peak value.  The same seven-bin constraint is used for this
fit.  The Gaussian mean and standard deviation are calculated from the
polynomial coefficients, and the symmetric fit results are accepted
unless the lower-half fit results in a smaller mean.  This histogram
and polynomial fit process is repeated again, with updated bin size
based on the previous iteration standard deviation, if the calculated
standard deviation is not larger than 75\% of the initial guess
(suggesting an issue with the initial bin size).

With this two-stage calculation performed across all subdivisions of
the mosaicked OTA image, and missing model pixels filled with the
average of their neighbors, the final background model is stored on
disk as a $13\times{}13$ image for the GPC1 chips with header entries
listing the binning used.  The full scale background image is then
constructed by bilinearly interpolating this binned model, and this is
subtracted from the science image.  Each object in the photometric
catalog has a SKY and SKY\_SIGMA value determined from the background
model mean and standard deviation.

Although this background modeling process works well for most of the
sky, astronomical sources that are large compared to the
$800\times{}800$ pixel subdivisions can bias the calculated background
level high, resulting in an oversubtraction near that object.  The
most common source that can cause this issue are large galaxies, which
can have their own features modeled as being part of the background.
For the specialized processing of M31, which covers an entire pointing
of GPC1, the measured background was added back to the \IPPstage{chip}
stage images, but this special processing was not used for the large
scale $3\pi$ PV3 reduction.

\subsection{Burntool / Persistence effect}
\label{sec:burntool}

Pixels that approach the saturation point on GPC1 (see
Section~\ref{sec:diffraction_spikes}) introduce ``persistent charge''
on that and subsequent images.  During the read out process of a cell
with such a bright pixel, some of the charge remains in the undepleted
region of the silicon and is not shifted down the detector column
toward the amplifier.  This charge remains in the starting pixel and
slowly leaks out of the undepleted region, contaminating subsequent
pixels during the read out process, resulting in a ``burn trail'' that
extends from the center of the bright source away from the amplifier
(vertically along the pixel columns toward the top of the cell).

This incomplete charge shifting in nearly full wells continues as each
row is read out.  This results in a remnant charge being deposited in
the pixels that the full well was shifted through.  In following
exposures, this remnant charge leaks out, resulting in a trail that
extends from the initial location of the bright source on the previous
image towards the amplifier (vertically down along the pixel column).
This remnant charge can remain on the detector for up to thirty
minutes.

Both of these types of persistence trails are measured and optionally
repaired via the \IPPprog{burntool} program.  This program does an
initial scan of the image, and identifies objects with pixel values
higher than a conservative threshold of 30000 DN.  The trail from the
peak of that object is fit with a one-dimensional power law in each
pixel column above the threshold, based on empirical evidence that
this is the functional form of this persistence effect.  This fit also
matches the expectation that a constant fraction of charge is
incompletely transferred at each shift beyond the persistence
threshold.  Once the fit is done, the model can be subtracted from
the image.  The location of the source is stored in a table along
with the exposure PONTIME, which denotes the number of seconds since
the detector was last powered on and provides an internally
consistent time scale.

\begin{figure}
  \centering
  \begin{minipage}{0.45\hsize}
    \includegraphics[width=0.9\hsize,angle=0,clip]{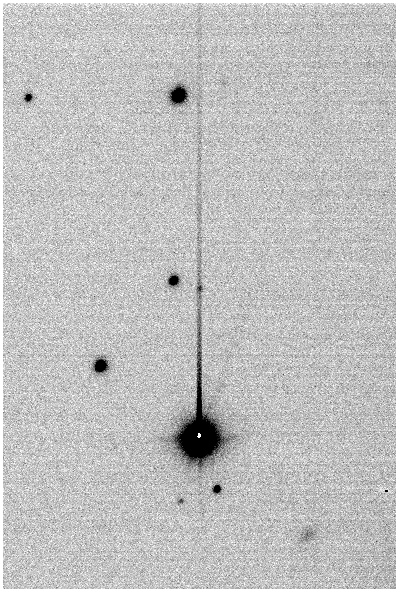}
  \end{minipage}%
  \begin{minipage}{0.45\hsize}
    \includegraphics[width=0.9\hsize,angle=0,clip]{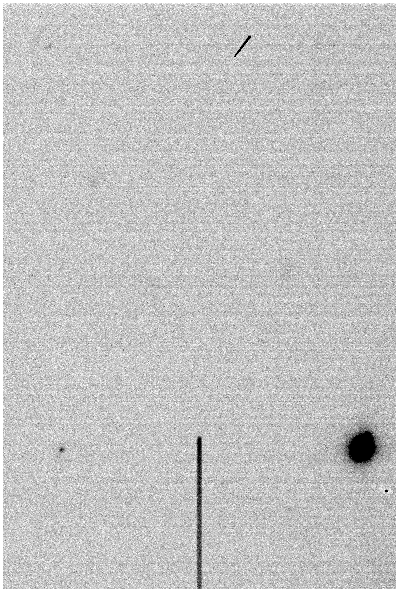}
  \end{minipage}
  \begin{minipage}{0.45\hsize}
    \includegraphics[width=0.9\hsize,angle=0,clip]{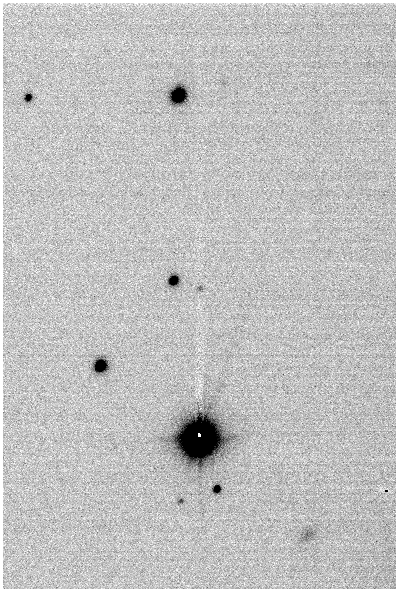}
  \end{minipage}%
  \begin{minipage}{0.45\hsize}
    \includegraphics[width=0.9\hsize,angle=0,clip]{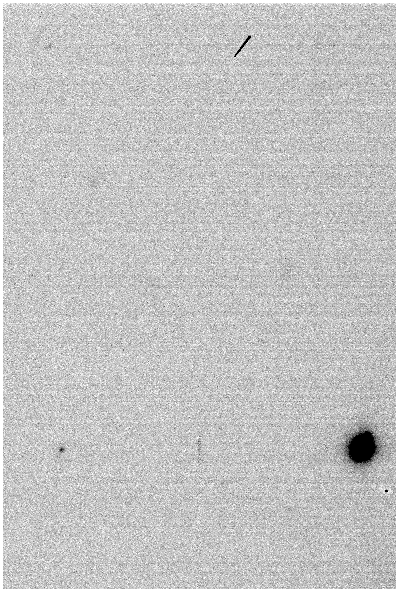}
  \end{minipage}
  \caption{{\bf Persistent Charge:}  Example of OTA11 cell xy50 on exposures o5677g0123o (left) and o5677g0124o (right).  The top panels show the image with all appropriate detrending steps, but without burntool, and the bottom show the same with burntool applied.  There is some slight over subtraction in fitting the initial trail, but the impact of the trail is greatly reduced in both exposures.}
  \label{fig:burntool images}
\end{figure}

\begin{figure}
  \centering
  \includegraphics[width=0.9\hsize,angle=0,clip]{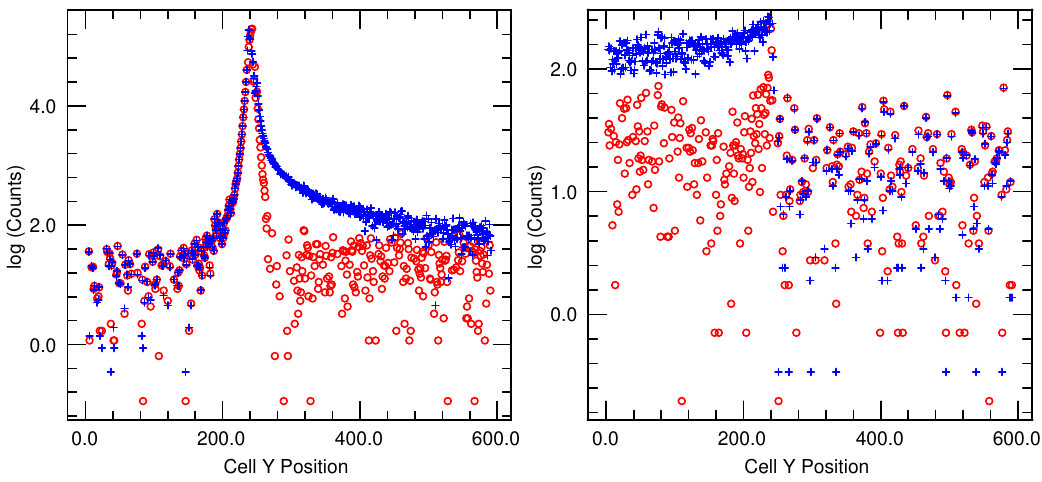}

  \caption{{\bf Burntool Correction:} Example of a profile cut along
    the y-axis through a bright star on exposure o5677g0123o OTA11 in
    cell xy50 (left panel) and on the subsequent exposure o5677g0124o
    (right panel).  In both figures, the blue pluses show the image
    corrected with all appropriate detrending steps, but without
    burntool applied, illustrating the amplitude of the persistence
    trails.  The red circles show the same data after the burntool
    correction, which reduces the impact of these features.  Both
    exposures are in the \gps{} filter with exposure times of 43s}

  \label{fig:burntool plot}
\end{figure}

For subsequent exposures, the table associated with the previous image
is read in, and after correcting trails from the stars on the new
image, the positions of the bright stars from the table are used to
check for remnant trails from previous exposures on the image.  These
are fit and subtracted using a one-dimensional exponential model,
again based on empirical studies.  The output table retains this
remnant position for 2000 seconds after the initial PONTIME recorded.
This allows fits to be attempted well beyond the nominal lifetime of
these trails.  Figure \ref{fig:burntool images} shows an example of a
cell with a persistence trail from a bright star, the post-correction
result, as well as the pre and post correction versions of the same
cell on the subsequent exposure.  The profiles along the detector
columns for these two exposures are presented in Figure
\ref{fig:burntool plot}.

Using this method of correcting the persistence trails has the
challenge that it is based on fits to the raw image data, which may
have other signal sources not determined by the persistence effect.
The presence of other stars or artifacts in the detector column can
result in a poor model to be fit, resulting in either an over- or
under-subtraction of the trail.  For this reason, the image mask is
marked with a value indicating that this correction has been applied.
These pixels are not fully excluded, but they are marked as suspect,
which allows them to be excluded from consideration in subsequent
stages, such as image stacking.

The cores of very bright stars can also be deformed by this process,
as the burntool fitting subtracts flux from only one side of the star.
As most stars that result in persistence trails already have saturated
cores, they are already ignored for the purpose of PSF determination
and are flagged as saturated by the photometry reduction.

\subsection{Non-linearity Correction}
\label{sec:nonlinearity}

The pixels of GPC1 are not uniformly linear at all flux levels.  In
particular, at low flux levels, some pixels have a tendency to sag
relative to the expected linear value.  This effect is most pronounced
along the edges of the detector cells, although some entire cells show
evidence of this effect.

To correct this sag, we studied the behavior of a series of flat
frames for a ramp of exposure times with approximate logarithmically
equal spacing between 0.01s and 57.04s.  As the exposure time
increases, the signal on each pixel also increases in what is expected
to be a linear manner.  Each of the flat exposures in this ramp is
overscan corrected, and then the median is calculated for each cell,
as well as for the rows and columns within ten pixels of the edge of
the science region.  From these median values at each exposure time
value, we can construct the expected trend by fitting a linear model
for the region considered.  This fitting was limited to only the range
of fluxes between 12000 and 38000 counts, as these ranges were found
to match the linear model well.  This range avoids the non-linearity
at low fluxes, as well as the possibility of high-flux non-linearity
effects.

% An example of this data is shown in Figure~\ref{fig: nonlinearity}.  

We store the average flux measurement and deviation from the linear
fit for each exposure time for each region on all detector cells in
the linearity detrend look-up tables.  When this correction is
applied to science data, these lookup tables are loaded, and a linear
interpolation is performed to determine the correction needed for the
flux in that pixel.  This look up is performed for both the row and
column of each pixel, to allow the edge correction to be applied where
applicable, and the full cell correction elsewhere.  The average of
these two values is then applied to the pixel value, reducing the
effects of pixel nonlinearity.

This non-linearity effect appears to be stable in time for the
majority of the detector pixels, with little evident change over the
survey duration.  However, as the non-linearity is most pronounced at
the edges of the detector cells, those are the regions where the
correction is most likely to be incomplete.  Because of this fact,
most pixels in the static mask with either the DARKMASK or FLATMASK
bit set are found along these edges.  As the non-linearity correction
is unable to reliably restore these pixels, they produce inconsistent
values after the dark and flat have been applied, and are therefore
rejected.

% this figure does not really clarify anything
% \begin{figure}
%   \centering
%   \includegraphics[width=0.9\hsize,angle=0,clip]{images/linearity_XY27_xy16.png}
%   \caption{Example of the linearity correction as a fraction of observed flux for OTA27, cell xy16.}
%   \label{fig: nonlinearity}
% \end{figure}

\subsection{Pattern correction}
\label{sec:pattern}

\subsubsection{Pattern Row}
\label{sec:pattern.row}
%% Statistics so I have them written down somewhere
%% chipProcessedImfile.bg/bg_stdev by filter for XY33 (a good chip)
%% filter  bg_mean stdev median Qsig                              bg_stdev_mean stdev median Qsig
%% g        36.37422026669   64.64175104057  32.693   6.10284     14.696938349131  78.80460307171  8.8401  0.5417843
%% r       200.96143304525  471.87743546238 117.105  94.55608     33.854672792146  79.01642728089 13.4564  5.3771355
%% i       447.00504994458  938.38517801037 286.810 154.71397     57.298335510188  99.38392923935 20.0217 24.2254723
%% z       317.54933679054  390.38930252748 241.014 114.13316     48.359069000176  94.44452756094 17.9404  9.1535209
%% y       371.09019536218  293.57439970375 288.481 133.38769     43.724342221691 135.04286534327 19.9029  7.5396461

As discussed above in the dark and noisemap sections, certain
detectors have significant bias offsets between adjacent rows, caused
by drifts in the bias level due to cross talk.  The magnitude of these
offsets increases as the distance from the readout amplifier and
overscan region increases, resulting in horizontal streaks that are
more pronounced along the large $x$ pixel edge of the cell.  As the
level of the offset is apparently random between exposures, the dark
correction cannot fully remove this structure from the images, and the
noisemap value only indicates the level of the average variance added
by these bias offsets.  Therefore, we apply the PATTERN.ROW correction
in an attempt to mitigate the offsets and correct the image values.
To force the rows to agree, a second order clipped polynomial is
fitted to each row in the cell.  Four fit iterations are run and
pixels $2.5\sigma$ deviant (chosen empirically) are excluded from
subsequent fits in order to minimize the bias from stars and other
astronomical sources in the pixels.  This final trend is then
subtracted from that row.  Simply doing this subtraction will also
have the effect of removing the background sky level.  To prevent
this, the constant and linear terms for each row are stored, and
linear fits are made to these parameters as a function of row,
perpendicular to the initial fits.  This produces a plane that is
added back to the image to restore the background offset and any
linear ramp that exists in the sky.

These row-by-row variations have the largest impact on data taken in
the \gps{} filter, as the read noise is the dominant noise source in
that filter.  At longer wavelengths, the noise from the Poissonian
variation in the sky level increases.  The PATTERN.ROW correction is
still applied to data taken in the other filters, as the increase in
sky noise does not fully obscure the row-by-row noise.

%% GPC1 tuning describe in email from Peter Onaka 2009.11.30,
%% with notes in GPC1TuningTestLog.pdf

This correction was required on all cells on all OTAs prior to
2009-12-01, at which point a modification of the camera clocking phase
delays reduced the scale of the row-by-row offsets for the majority of
the OTAs.  As a result, we only apply this correction to the cells
where it is still necessary, as shown in Figure \ref{fig: pattern row
  cells}.  A list of these cells is in Table
\ref{tab:pattern_row_cells}.

\begin{figure}
  \centering
  \begin{minipage}{0.45\hsize}
    \includegraphics[width=0.9\hsize,angle=0,clip]{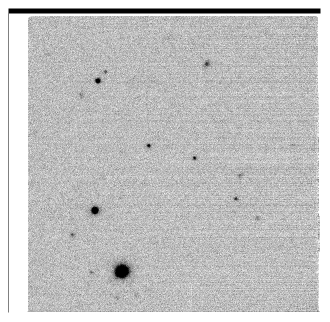}
  \end{minipage}%
  \begin{minipage}{0.45\hsize}
    \includegraphics[width=0.9\hsize,angle=0,clip]{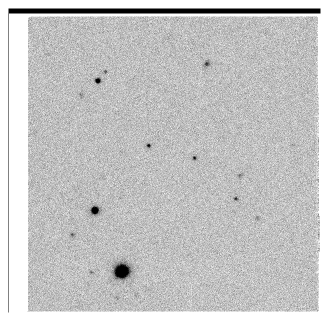}
  \end{minipage}
  \caption{{\bf Correlated Noise:} Example of the PATTERN.ROW correction on exposure o5379g0103o OTA57 cell xy01 (\ips{} filter 45s).  The left panel shows the cell with all appropriate detrending except the PATTERN.ROW, and the right shows the same cell with PATTERN.ROW applied.  The correction reduces the correlated noise on the right side, which is most distant from the read out amplifier.  There is a slight over subtraction along the rows near the bright star.}
  \label{fig: pattern row example}
\end{figure}

Although this correction largely resolves the row-by-row offset issue
in a satisfactory way, large and bright astronomical objects can bias
the fit significantly.  This results in an oversubtraction of the
offset near these objects.  As the offsets are calculated on the pixel
rows, this oversubtraction is not uniform around the object, but is
preferentially along the horizontal x axis of the object.  Most
astronomical objects are not significantly distorted by this, with
this only becoming on issue for only bright objects comparable to the
size of the cell (598 pixels = 150").  Figure \ref{fig: pattern row example} 
shows an example of a cell pre- and post-correction.

\begin{deluxetable}{lcccc}
  \tablecolumns{3}
  \tablewidth{0pc}
  \tablecaption{Cells which have PATTERN.ROW correction applied}
  \tablehead{\colhead{OTA} & \colhead{Cell columns} & \colhead{Additional cells}}
  \startdata
  OTA11 &  & xy02, xy03, xy04, xy07 \\
  OTA14 &  & xy23 \\
  OTA15 & 0 & \\
  OTA27 & 0, 1, 2, 3, 7 & \\
  OTA31 & 7 & \\
  OTA32 & 3, 7 & \\
  OTA45 & 3, 7 & \\
  OTA47 & 0, 3, 5, 7 & \\
  OTA57 & 0, 1, 2, 6, 7 & \\
  OTA60 &  & xy55 \\
  OTA74 & 2, 7 & \\
  \enddata
  \label{tab:pattern_row_cells}
\end{deluxetable}

\begin{figure}
  \centering
  \includegraphics[width=0.9\hsize,angle=0,clip]{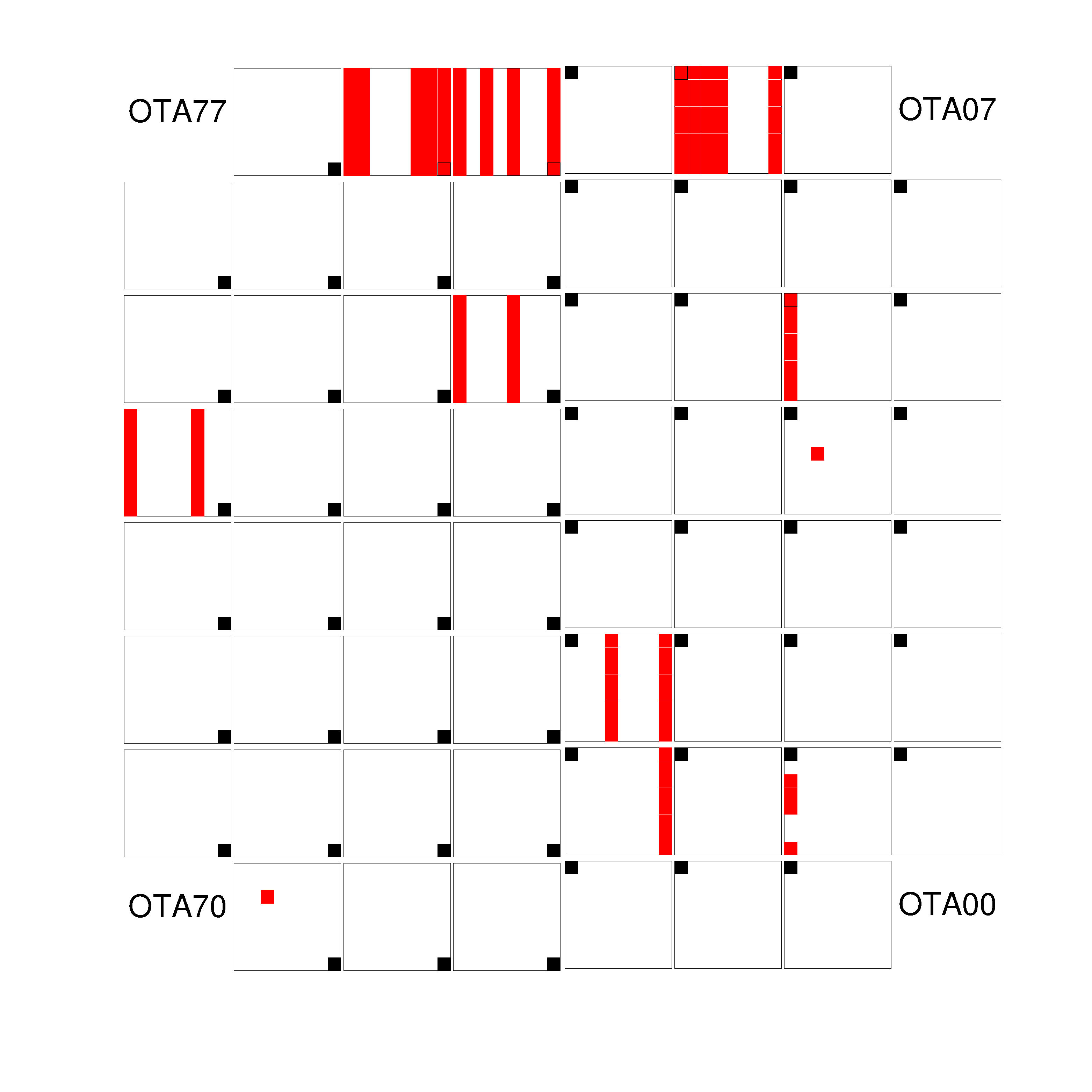}
  \caption{Diagram illustrating in red which cells on GPC1 require the PATTERN.ROW correction to be applied.  The footprint of each OTA is outlined, and cell xy00 is marked with either a filled box or an outline.  The labeling of the non-existent corner OTAs is provided to orient the focal plane.}
  \label{fig: pattern row cells}
\end{figure}

\subsubsection{Pattern Continuity}

The background sky levels of cells on a single OTA do not always have
the same value.  Despite having dark and flat corrections applied,
adjacent cells may not match even for images of nominally empty sky.
In addition, studies of the background level indicate that the
row-by-row bias can introduce small background gradient variations
along the rows of the cells that are not stable.  This common feature
across the columns of cells results in a ``saw tooth'' pattern
horizontally across an the mosaicked OTA, and as the background model
fits a smooth sky level, this induces over- and under subtraction at
the cell boundaries.

The PATTERN.CONTINUITY correction, attempts to match the edges of a
cell to those of its neighbors.  For each cell, a thin box 10 pixels
wide running the full length of each edge is extracted and the median
of unmasked values is calculated for that box.  These median values
are then used to construct a vector of the sum of the differences
between that cell's edges and the corresponding edge on any adjacent
cell $\Delta$.  A matrix $A$ of these associations is also
constructed, with the diagonal containing the number of cells adjacent
to that cell, and the off-diagonal values being set to -1 for each
pair of adjacent cells.  The offsets needed for each chip, $\zeta$ can
then be found by solving the system $A \zeta = \Delta$. A cell with the
maximum number of neighbors, usually cell xy11, the first cell not on
the edge of the OTA, is used to constrain the system, ensuring that
that cell has zero correction and that there is a single solution.

For OTAs that initially show the saw tooth pattern, the effect of this
correction is to align the cells into a single ramp, at the expense of
the absolute background level.  However, as we subtract off a smooth
background model prior to doing photometry, these deviations from an
absolute sky level do not affect photometry for point sources and
extended sources smaller than a single cell.  The fact that the
final ramp is smoother than it would be otherwise also allows for the
background subtracted image to more closely match the astronomical
sky, without significant errors at cell boundaries.  An example of the
effect of this correction on an image profile is shown in Figure
\ref{fig:dark switching}.

\section{GPC1 Detrend Construction}
\label{sec:detrend construction}

The various master detrend images for GPC1 are constructed using a
common approach.  A series of appropriate exposures is selected from
the database, and processed with the \IPPprog{ppImage} program, which
is designed to do multiple image processing operations.  The
processing steps applied to the images depend on the type of master
detrend to be constructed.  In general, the input exposures to the
detrend have all prior stages of detrend processing applied.  Table
\ref{tab:detrend ppImage} summarizes stages applied for the detrends
we construct.

Once the input data has been prepared, the \IPPprog{ppMerge} program
is used to combine the inputs.  In some cases, this is the
mathematical average, but in other cases it is a fit across the
inputs.  Table \ref{tab:detrend ppMerge} lists some of the properties
of the process for the detrends, including how discrepant values are
removed and the combination method used.  The outputs from this step
have the format of the detrend under construction.  After
construction, these combined outputs are applied to the processed
input data.  This creates a set of residual files that are checked to
determine if the newly created detrend correctly removes the detector
dependent signal.

This process of detrend construction and testing can be iterated, with
individual exposures excluded if they are found to be contaminating
the output.  The construction of detrends is largely automatic, but
manual intervention is needed to accept the detrend for use on science
data.  If the final detrend has sufficiently small residuals, then the
iterations are stopped and the detrend is finalized by selecting the
date range to which it applies.  This allows subsequent science
processing to select the detrends needed based on the observation
date.  Table \ref{tab:detrend list} lists the set of detrends used in
the PV3 processing.

\begin{deluxetable*}{lcccc}
  \tablecolumns{5}
  \tablewidth{0pc}
  \tablecaption{Detrend Construction Processing}
  \tablehead{\colhead{Detrend Type} & \colhead{Overscan Subtracted} & \colhead{Nonlinearity Correction} & \colhead{Dark Subtracted} & \colhead{Flat Applied} }
  \startdata
  LINEARITY & Y & & & \\
%%  DARKMASK  & Y & Y & Y & \\
%%  FLATMASK  & Y & Y & Y & Y \\
%%  CTEMASK   & Y & Y & Y & Y \\
  DARK      & Y & Y & & \\
%%  NOISEMAP  & Y & Y & & \\
  FLAT      & Y & Y & Y & \\
  FRINGE    & Y & Y & Y & Y \\
  DARKMASK  & Y & Y & Y & \\
  FLATMASK  & Y & Y & Y & Y \\
  CTEMASK   & Y & Y & Y & Y \\
  NOISEMAP  & Y & Y & & \\
  \enddata
  \label{tab:detrend ppImage}
\end{deluxetable*}

\begin{deluxetable*}{lcccc}
  \tablecolumns{5}
  \tablewidth{0pc}
  \tablecaption{Detrend Merge Options}
  \tablehead{\colhead{Detrend Type} & \colhead{Iterations} & \colhead{Threshold} & \colhead{Additional Clipping} & \colhead{Combination Method} }
  \startdata
  DARKMASK  & 3 & $8\sigma$ & & Mask if $>10\%$ rejected \\
  FLATMASK  & 3 & $3\sigma$ & & Mask if $>10\%$ rejected \\
  CTEMASK   & 2 & $2\sigma$ & & Clipped mean; mask if $\sigma^2/\langle I\rangle < 0.5$ \\
  DARK      & 2 & $3\sigma$ & & Clipped mean \\
  NOISEMAP  & 2 & $3\sigma$ & & Mean \\
  FLAT      & 1 & $3\sigma$ & Top $30\%$; Bottom $10\%$ & Mean \\
  FRINGE    & 2 & $3\sigma$ & & Clipped mean \\
  \enddata
  \label{tab:detrend ppMerge}
\end{deluxetable*}

\begin{deluxetable*}{lclll}
  \tablecolumns{5}
  \tablewidth{0pc}
  \tablecaption{PV3 Detrends}
  \tablehead{\colhead{Detrend Type} & \colhead{Detrend ID} &
    \colhead{Start Date (UT)} & \colhead{End Date (UT)} & \colhead{Note} }
  \startdata
  LINEARITY & 421  & 2009-01-01 00:00:00 & & \\
  MASK      & 945  & 2009-01-01 00:00:00 & & \\
            & 946  & 2009-12-09 00:00:00 & & \\
            & 947  & 2010-01-01 00:00:00 & & \\
            & 948  & 2011-01-06 00:00:00 & & \\
            & 949  & 2011-03-09 00:00:00 & 2011-03-10 23:59:59 & \\
            & 950  & 2011-08-02 00:00:00 & & \\
            & 1072 & 2015-12-17 00:00:00 & & Update OTA62 mask \\
  DARK      & 223  & 2009-01-01 00:00:00 & 2009-12-09 00:00:00 & \\
            & 229  & 2009-12-09 00:00:00 & & \\
            & 863  & 2010-01-23 00:00:00 & 2011-05-01 00:00:00 & A-mode \\
            & 864  & 2011-05-01 00:00:00 & 2011-08-01 00:00:00 & \\
            & 865  & 2011-08-01 00:00:00 & 2011-11-01 00:00:00 & \\
            & 866  & 2011-11-01 00:00:00 & 2019-04-01 00:00:00 & \\
            & 869-935 & 2010-01-25 00:00:00\tablenotemark{a} & 2011-04-25 23:59:59\tablenotemark{a} & B-mode \\
  VIDEODARK & 976  & 2009-01-01 00:00:00 & 2009-12-09 00:00:00 & \\
            & 977  & 2009-12-09 00:00:00 & 2010-01-23 00:00:00 & \\
            & 978  & 2010-01-23 00:00:00 & 2011-05-01 00:00:00 & A-mode \\
            & 979  & 2011-05-01 00:00:00 & 2011-08-01 00:00:00 & \\
            & 980  & 2011-08-01 00:00:00 & 2011-11-01 00:00:00 & \\
            & 981  & 2011-11-01 00:00:00 & 2019-04-01 00:00:00 & \\
            & 982-1048 & 2010-01-25 00:00:00\tablenotemark{a} & 2011-04-25 23:59:59\tablenotemark{a} & B-mode \\
            & 1049 & 2010-09-12 00:00:00 & 2011-05-01 00:00:00 & A-mode with OTA47fix \\
  NOISEMAP  & 963  & 2008-01-01 00:00:00 & 2010-09-01 00:00:00 & \\
            & 964  & 2010-09-01 00:00:00 & 2011-05-01 00:00:00 & \\
            & 965  & 2011-05-01 00:00:00 & & \\
  FLAT      & 300  & 2009-12-09 00:00:00 & & \gps{} filter \\
            & 301  & 2009-12-09 00:00:00 & & \rps{} filter \\ 
            & 302  & 2009-12-09 00:00:00 & & \ips{} filter \\
            & 303  & 2009-12-09 00:00:00 & & \zps{} filter \\
            & 304  & 2009-12-09 00:00:00 & & \yps{} filter \\
            & 305  & 2009-12-09 00:00:00 & & \wps{} filter \\
  FRINGE    & 296  & 2009-12-09 00:00:00 & & \\
  ASTROM    & 1064 & 2008-05-06 00:00:00 & & \\
  \enddata
  \tablenotetext{a}{These dates mark the beginning and ending of the two-mode dark models, between which multiple dates use the B-mode dark.}
  \label{tab:detrend list}
\end{deluxetable*}

\section{Warping}
\label{sec:warping}

In order to perform image combination operations (stacking and
differences), the individual OTA images are geometrically transformed
to a set of images with a consistent and uniform relationship between
sky coordinates and image pixels.  This warping operation transforms
the image pixels from the regular grid laid out on the chips in the
camera to a system of pixels with consistent geometry for a location
on the sky.

The new image coordinate system is defined by one of a number of
``tessellations'' which specify how the sky is divided into individual
images.  A single tessellation starts with a collection of projection
centers distributed across the sky.  A grid of image pixels about each
projection center corresponds to sky positions via a projection with a
specified pixel scale and rotation.  In general, the pixel grid within
the projection is defined as a simplified grid with the y-axis aligned
to the Declination lines and no distortion terms.  The projection
centers are typically separated by several degrees on the sky; for
pixel scales appropriate to GPC1, the resulting collection of pixels
would be unwieldy in terms of memory in the processing computer.  The
pixel grid is thus subdivided into smaller sub-images called
'skycells'.

A tessellation can be defined for a limited region, with only a small
number of projection centers (e.g., for processing the M31 region), or
even a single projection center (e.g., for the Medium Deep fields).
For the $3\pi$ survey, the tessellation contains projection centers
covering the entire sky.  The version used to for the PV3 analysis is
called the \ippmisc{RINGS.V3}.  This tessellation consists of 2643
projection centers spaced every four degrees in DEC, with RA spacing
of approximately four degrees, adjusted to ensure an integer number of
equal-sized regions.  \ippmisc{RINGS.V3} uses a pixel scale of
$0\farcs{}25$ per pixel.  The projections subdivided into a
$10\times{}10$ grid of skycells, with an overlap region of
60\arcsec\ between adjacent skycells to ensure that objects of modest
size are not split on all images.  The coordinate system used for
these images matches the parity of the sky, with north in the positive
$y$ direction and east to the negative $x$ direction.

After the detrending and photometry, the detection catalog for the
full camera is fit to the reference catalog, producing astrometric
solutions that map the detector focal plane to the sky, and map the
individual OTA pixels to the detector focal plane
\citep[see][]{magnier2017.calibration}.  This solution is then used to
determine which skycells the exposure OTAs overlap.

For each output skycell, all overlapping OTAs and the calibrated
catalog are read into the \IPPprog{pswarp} program.  The output warp
image is broken into $128\times{}128$ pixel grid boxes.  For purposes
of speed, each grid box has a locally linear map calculated that
converts the output warp image coordinates to the input chip image
coordinates.  By doing the transformation in this direction, each
output pixel has a unique sampling position on the input image
(although it may be off the image frame and therefore not populated),
guaranteing that all output pixels are addressed, and thus preventing
gaps in the output image due to the spacing of the input pixels.

With the locally linear grid defined, Lanczos interpolation
\citep{lanczos1956applied} with filter size parameter $a = 3$ on the
input image is used to determine the values to assign to the output
pixel location.  This interpolation kernel was chosen as a compromise
between simple interpolations and higher-order Lanczos kernels, with
the goal of limiting the smear in the output image while avoiding
the high-frequency ringing generated by higher order kernels.  This
process is repeated for all grid boxes, for all input images, and for
each output image product: the science image, the variance, and the
mask.  The image values are scaled by the absolute value of the
Jacobian determinant of the transformation for each grid box.  This
corrects the pixel values for the possible change in pixel area due to
the transformation.  Similarly, the variance image is scaled by the
square of this value, again to correctly account for the pixel area
change.

The interpolation constructs the output pixels from more than one
input pixel, which introduces covariance between pixels.  For each
locally-linear grid box, the covariance matrix is calculated from the
kernel in the center of the 128 pixel range.  Once the image has been
fully populated, this set of individual covariance matrices are
averaged to create the final covariance for the full image.

An output catalog is also constructed from the full exposure input
catalog, including only those objects that fall on the new warped image.
These detections are transformed to match the new image location, and
to scale the position uncertainties based on the new orientation.

The output image also contains header keywords SRC\_nnnn, SEC\_nnnn,
MPX\_nnnn, and MPY\_nnnn that define the mappings from the warped
pixel space to the input images.  The 'nnnn' for each keyword has the
values 0000, 0001, etc., up to the number of input images.  The SRC
keyword lists the input OTA name, and the SEC keyword lists the image
section that the mapping covers.  The MPX and MPY contain the
back-transformation linearized across the full chip.  These parameters
are stored in a string listing the reference position in the chip
coordinate frame, the slope of the relation in the warp $x$ axis, and
the slope of the relation in the warp $y$ axis.  From these keywords,
any position in the warp can be mapped back to the location in any of
the input OTA images, with some reduction in accuracy.

\begin{figure}
  \centering
  \includegraphics[width=0.9\hsize,angle=0,clip]{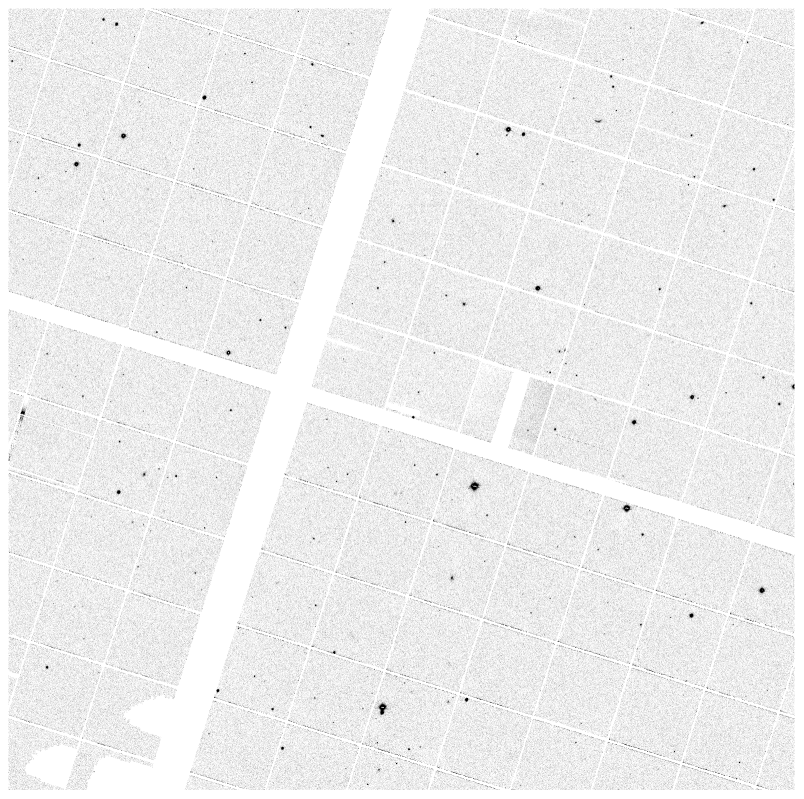}
  \caption{Example of the warp image for skycell skycell.1146.095
    centered at ($\alpha,\delta$) = (11.934, -4.197) for exposure
    o5104g0266o, (2009-09-30, 60s \rps{} filter).  The data from four
    OTAs contribute to this image, although they are all truncated by
    the skycell boundaries.  This skycell image is aligned such that
    north points to the top of the image, and east to the left.  The
    contributing OTAs are OTA20, OTA21, OTA30, OTA31.}
  \label{fig:warp image}
\end{figure}

\begin{figure}
  \centering
  \includegraphics[width=0.9\hsize,angle=0,clip]{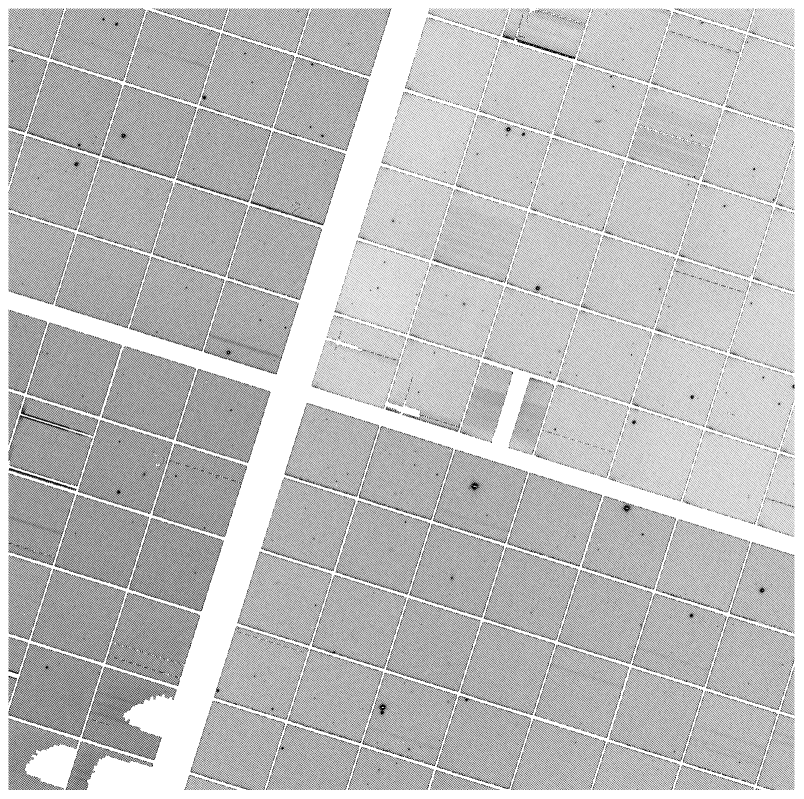}
  \caption{Example of the warp variance image for skycell
    skycell.1146.095 of exposure o5104g0266o, the same as in Figure
    \ref{fig:warp image}.  This variance map retains information about
    the higher flux levels that were found in burntool corrected
    persistence trails, which appear here as streaks along the
    original OTA y axis.  The dark glows that are corrected in the
    dark model are also more visible, especially on certain cell
    edges.  As both of these effects are corrected in the science
    image, there are no significant features visible there.}
  \label{fig:warp variance}
\end{figure}

\begin{figure}
  \centering
  \includegraphics[width=0.9\hsize,angle=0,clip]{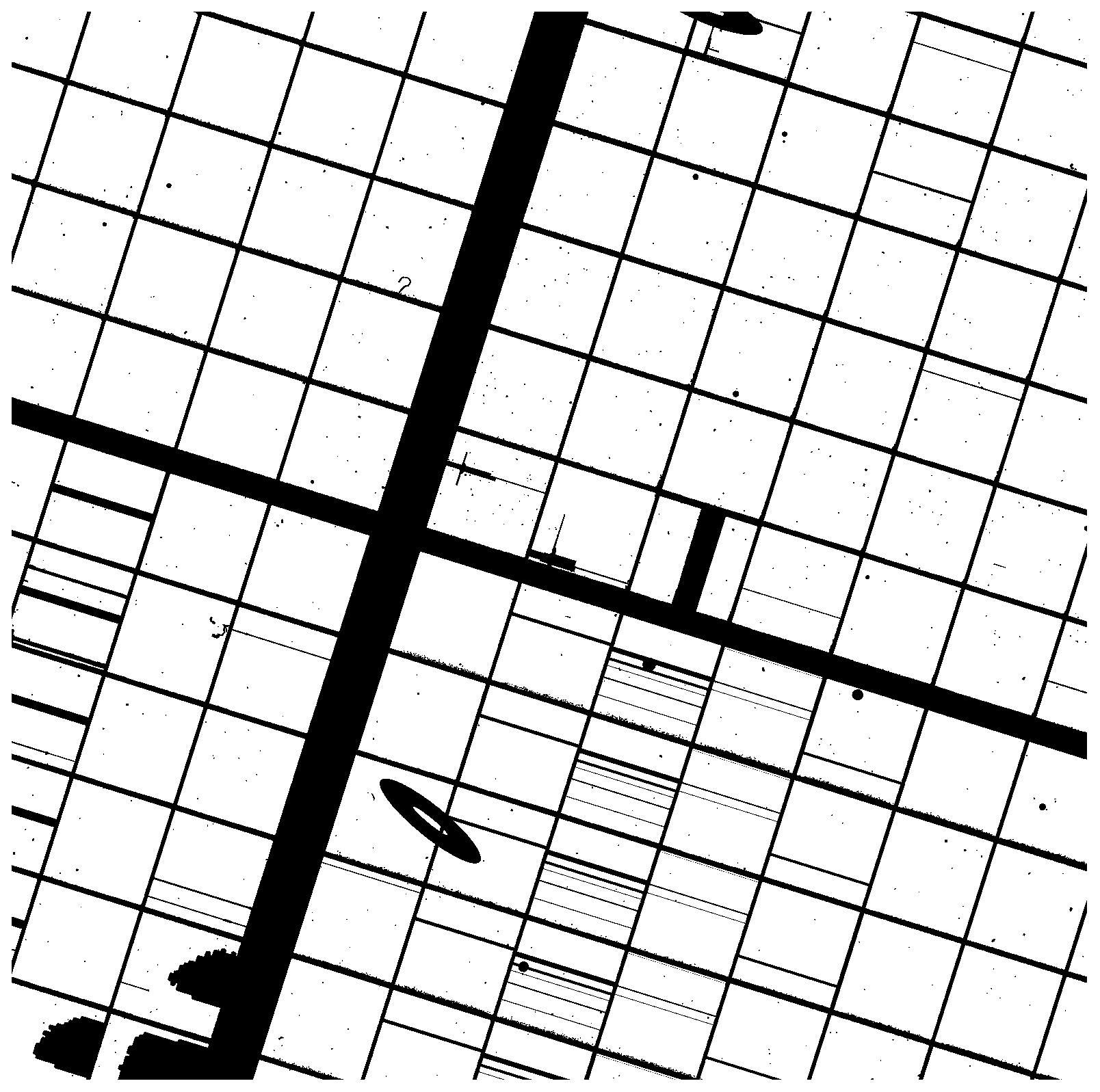}
  \caption{Example of the warp mask image for skycell skycell.1146.095
    of exposure o5104g0266o, the same as in Figure \ref{fig:warp
      image}.  This mask image shows the many small defects removed
    from the image, along with larger advisory trails on corrected
    burntool trails.  The saturated cores of the bright stars are also
    masked, along with the diffraction spikes found on these stars.  A
    ghost mask is visible just below the center as an elliptical
    region.
%    In addition OTA24 shows the precautionary crosstalk bleed masks
%    for the two brightest stars applied to all cells within the same
%    row.
  \label{fig:warp mask}
  }
\end{figure}

Examples of a warped signal, variance, and mask image are illustrated
in Figures~\ref{fig:warp image} through \ref{fig:warp mask}.

\section{Stacking}
\label{sec:stacking}

Once individual exposures have been warped onto a common projection
system, they can be combined pixel-by-pixel regardless of their
original orientation.  Creating a stacked image by co-adding the
individual warps increases the signal to noise, allowing for the
detection of objects that would not be sufficiently significant to be
measured from a single image.  Creating this stack also allows a more
complete image to be constructed that has fewer regions masked due to
the gaps between cells and OTAs.  This deeper and more complete image
can also be used as a template for subtraction to find transient
sources.

As part of the stacking process, the collection of input pixels for a
given output stack pixel are checked for consistency and outliers are
rejected.  Varying image quality makes a pixel-by-pixel check for
outliers challenging in the vicinity of brighter stars.  Pixels in the
wings of bright stars are liable to be over-rejected as the image
quality changes because the flux observed at a given position varies
as its location on the stellar profile changes.  To avoid this effect,
we convolve all input images to a common PSF before making the
pixel-by-pixel comparison.  This PSF-matching technique allows us to
detect inconsistent pixels even in the sensitive wings of bright objects.

For the $3\pi$ survey, the stacked image is comprised of all warp
frames for a given skycell in a single filter.  The source catalogs
and image components are loaded into the \IPPprog{ppStack} program to
prepare the inputs and stack the frames.

Once all files are ingested, the first step is to measure the size and
shapes of the input image PSFs.  We exclude images that have a PSF
FWHM greater than 10 pixels (2.5 arcseconds), as those images have the
seeing far worse than average, and would degrade the final output
stack.  For the PV3 processing of the $3\pi$ survey, this size represents a PSF larger
than the $97$th percentile in all filters.  A target PSF for the stack
is constructed by finding the maximum envelope of all input PSFs,
which sets the target PSF to the largest value among the input PSFs
for a given position from the peak.  This PSF is then circularized to
ensure azimuthal symmetry, which prevents deconvolution of any of the
input images when matched to the target.

The input image fluxes are normalized to prevent differences in seeing
and sky transparency from causing discrepancies during pixel
rejection.  From the reference catalog calibrated input catalogs, we
have the instrumental magnitudes of all sources, along with the
airmass, image exposure time, and zeropoint.  All output stacks are
constructed to a target zeropoint of 25.0 in all filters, and to have
an airmass of 1.0.  The target zeropoint is arbitrary; 25.0 was chosen
to be roughly consistent with the PS1 zero points, while still being a
simple number.  The output exposure time is set to the sum of the
input exposure times, {\em regardless of whether those inputs are rejected
later in the combination process}.  We can determine the relative
transparency for each input image by comparing the magnitudes of
matched sources between the different images.  Each image then has a
normalization factor defined, equal to $\mathrm{norm}_{input} =
(ZP_\mathrm{input} - ZP_\mathrm{target}) -
\mathrm{transparency}_\mathrm{input} - 2.5 \times \log_{10}
(t_\mathrm{target} / t_\mathrm{input}) - \mathrm{F}_\mathrm{airmass} \times
(\mathrm{airmass}_\mathrm{input} - \mathrm{airmass}_\mathrm{target})$.
For the PV3 processing, the airmass factor
$\mathrm{F}_\mathrm{airmass}$ was set to zero, such that all flux
differences from differing exposure airmasses are assumed to be
included in the zeropoint and transparency values.

The zeropoint calibration performed here uses the calibration of the
individual input exposures against the reference catalog.  Upon the
conclusion of the survey, the entire set of detection catalogs is
further re-calibrated in the ``ubercal'' process \citep{2012ApJ...756..158S}.
This produces a more consistent calibration of each exposure across
the entire region of the sky imaged.  This further calibration is not
available at the time of stacking, and so there may be small residuals
in the transparency values as a result of this \citep{magnier2017.calibration}.

With the flux normalization factors and target PSF chosen, the
convolution kernels can be calculated for each image.  To calculate
the convolution kernels, we use the algorithm described by
\cite{1998ApJ...503..325A} and extended by \cite{2000AAS..144..363A}
to perform optimal image subtraction.  These `ISIS' kernels
\citep[named after the software package described
  by][]{1998ApJ...503..325A} are used with FWHM values of 1.5, 3.0,
and 6.0 pixels and polynomial orders of 6, 4, and 2.  Regions around
the sources identified in the input images are extracted, convolved
with the kernel, and the residual with the target PSF used to update
the parameters of the kernel via least squares optimization.  Stamps
that significantly deviate are rejected, although the squared residual
difference will increase with increasing source flux.  To mitigate
this effect, a parabola is fit to the distribution of squared
residuals as a function of source flux.  Stamps that deviate from this
fit by more than $2.5\sigma$ are rejected, and not used on further
kernel fit iterations.  This process is repeated twice, and the final
convolution kernel is returned.

This convolution may change the image flux scaling, so the kernel is
normalized to account for this.  The normalization factor is equal to
the ratio of $10^{-0.4 \mathrm{norm}_{input}}$ to the sum of the
kernel.  The image is multiplied by this factor, and the variance by
the square of it, scaling all inputs to the common zeropoint.

Once the convolution kernels are defined for each image, they are used
to convolve the image to match the target PSF.  Any input image that
has a kernel match $\chi^2$ value (defined as the sum of the RMS error
across the kernel) 4.0$\sigma$ or larger than the median
value is rejected from the stack.  Each image also has a weight
assigned, based on the image variance after convolution.  A full image
weight is then calculated for each input, with the weight,
$W_\mathrm{input}$ equal to the inverse of the median of the image
variance multiplied by the peak of the image covariance (from the
warping process).  This ensures that low signal-to-noise images are
down-weighted in the final combination.

Following the convolution, an initial stack is constructed.  For a
given pixel coordinate, the values at that coordinate are extracted
from all input images, with pixels masked excluded from consideration.
Images that only have a suspect mask bit (including the SUSPECT,
BURNTOOL, SPIKE, STREAK, STARCORE, and CONV.POOR bit values) are
appended to a suspect pixel list for preferential exclusion.
Following this, the pixel values are combined and tested to attempt to
identify discrepant input values that should be excluded.

If only a single input is available, the initial stack contains the
value from that single input.  If there are only two inputs, the
average of the two is used.  These cases are expected to occur only rarely in
the $3\pi$ survey, as there are many input exposures that overlap each
point on the sky.  For the more common case of three or more inputs, a
weighted average from the inputs is used, with the weight for each
image as defined above used for all pixels from that input image.
This weight is used for both the image and the exposure weighted
image:

\begin{eqnarray}
  \mathrm{Stack}_\mathrm{value} &=& \sum_i\left(\mathrm{value}_\mathrm{input} \times W_\mathrm{input}\right) / \sum_\mathrm{inputs}\left(W_\mathrm{input}\right) \\
  \mathrm{Stack}_\mathrm{exp weight} &=& \sum_i \left(\mathrm{exptime}_\mathrm{input} \times W_\mathrm{input}\right) / \sum_\mathrm{inputs}\left(W_\mathrm{input}\right) \\
\end{eqnarray}

The pixel exposure time is simply the sum of the input exposure time values, and the output variance is 

\begin{eqnarray}
  \mathrm{Stack}_\mathrm{variance} &=& 1 / \sum_i \left( 1 / \sigma^2_\mathrm{input} \right)
\end{eqnarray}

The output mask value is taken to be zero (no masked bits), unless
there were no valid inputs, in which case the BLANK mask bit is set.

Due to uncorrected artifacts that can occur on GPC1, and the fact that
they may not be fully masked to ensure all bad pixels are removed, it
is expected that some of the inputs for a given stack pixel are not in
agreement with the others.  In general, there is the population of
input pixel values around the correct astronomical level, as well as
possible populations at lower pixel value (such as due to an
over-subtracted burntool trail) and at higher pixel values (such as
that caused by an incompletely masked optical ghost).  Due to the
observation strategy to observe a given field twice to allow for
warp-warp difference images to be constructed to identify transient
detections, higher pixel values that come from sources like optical
ghosts that depend on the telescope pointing will come in pairs.
Detector artifacts will  appear in pairs as well.  The higher
pixel value contaminants are also potentially problematic as they may
appear to be real sources, prompting photometry to be performed on
false objects.  Because of the expectation that there are more positive
deviations than negative ones, there is a slight preference to  reject
higher pixel value outliers than lower pixel values, as described below.

Following the initial combination, a ``testing'' loop iterates in an
attempt to identify outlier points.  Again, if only one input is
available, that input is accepted.  If there are two inputs, $A$ and
$B$, then a check is made to see if $(0.5 \times (\mathrm{value}_A -
\mathrm{value}_B))^2 > 16 \times (\sigma^2_A + \sigma^2_B
+ (0.1 \times \mathrm{value}_A)^2 + (0.1 \times \mathrm{value}_B)^2)$, such that
the deviation of the inputs from their mean position is greater than
four times the sum of their measured uncertainties and a 10\%
systematic error term.  If this is the case, neither input is trusted,
and both are flagged for rejection

If the number of input pixels is larger than 6, then a Gaussian mixture
model analysis is run on the inputs fitting two sub populations, to
determine the likelihood that the distribution is best described by
an uni-modal model.  If this probability is less than $5\%$, then the
mean is taken from the bimodal sub population with the largest
fraction of inputs, as this should exclude any sub population
comprised of high pixel value outliers.

If the unimodal probability is greater than $5\%$ (indicating the
distribution is likely to be unimodal), or if there are insufficient
inputs for this mixture model analysis, the input values are passed to
an ``Olympic'' weighted mean calculation (both the lowest and highest
values are ignored in calculating the weighted mean).  We reject
$20\%$ of the number of inputs through this process.  The number of
bad inputs is set to $N_\mathrm{bad} = 0.2 \times N_\mathrm{input} +
0.5$, with the 0.5 term ensuring at least one input is rejected.  This
number is further separated into the number of low values to exclude,
$N_\mathrm{low} = N_\mathrm{bad} / 2$, which will default to zero if
there are few inputs, and $N_\mathrm{high} = N_\mathrm{low} -
N_\mathrm{bad}$.  After sorting the input values to determine which
values fall into the low and high groups, the remaining input values
are used in a weighted mean using the image weights above.

A systematic variance term is necessary to correctly scale how
discrepant points can be from the ensemble mean.  If the mixture model
analysis has been run, the Gaussian sigma from the largest sub
population is squared and used.  Otherwise, a $10\%$ systematic error
on the input values is used.  Each point then has a limit calculated
using a $4\sigma$ rejection

\begin{eqnarray}
  \mathrm{limit}_\mathrm{mixture\ model} &=& 4^2 \times (\sigma^2_\mathrm{input} + \sigma_\mathrm{mixture\ model}^2) \\
  \mathrm{limit}_\mathrm{default} &=& 4^2 \times (\sigma^2_\mathrm{input} + (0.1 \times \mathrm{value}_\mathrm{input})^2)
\end{eqnarray}

Each input pixel is then compared against this limit, and the most
discrepant pixel that has $(\mathrm{value}_\mathrm{input} -
\mathrm{mean})^2$ exceeding this limit is identified.  If there are
suspect pixels in the set, those pixels are marked for rejection,
otherwise this worst pixel is marked for rejection.  Following this step,
the combine and test loop is repeated until no more pixels are
rejected, up to a maximum number of iterations equal to $50\%$ of the
number of inputs.

With the initial list of rejected pixels generated, a rejection mask
is made for the input warp by constructing an empty image that has the
rejected pixels from that input set to a value of 1.0.  This image is
then convolved with a 5 pixel FWHM zeroth-order ISIS kernel.  Any
pixels that are above the threshold of 0.5 after this mask convolution
are marked as bad and will be rejected in the final combination.  If
more than 10\% of all pixels from an input image are rejected, then
the entire image is rejected as it likely has some systematic issue.

\begin{figure}
  \centering
  \includegraphics[width=0.9\hsize,angle=0,clip]{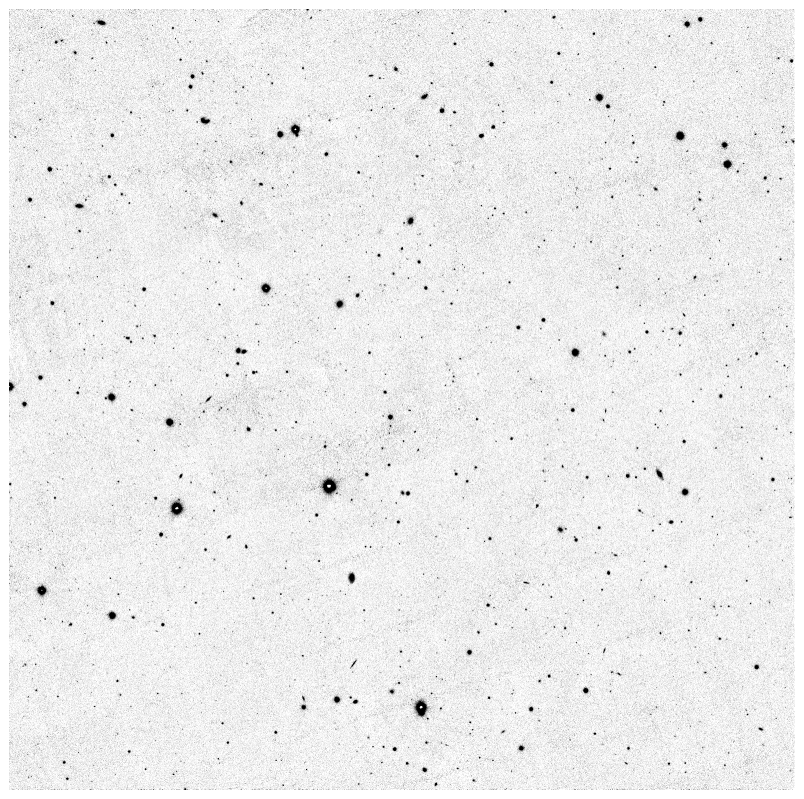}
  \caption{Example of the stack image for skycell skycell.1146.095
    centered at ($\alpha,\delta$) = (11.934, -4.197) in the \rps{}
    filter, stack\_id 3956997.  This stack includes 39 input images
    including o5104g0266o, the warp image in Figure \ref{fig:warp
      image}, and has a combined exposure time of 1880s.  Combining
    such a large number of input images removes the inter-cell and
    inter-chip gaps, providing a fully populated image.  In addition,
    the combined signal allows many more faint objects to be found
    than were visible on the single frame warp image.}

  \label{fig:stack image}
\end{figure}

\begin{figure}
  \centering
  \includegraphics[width=0.9\hsize,angle=0,clip]{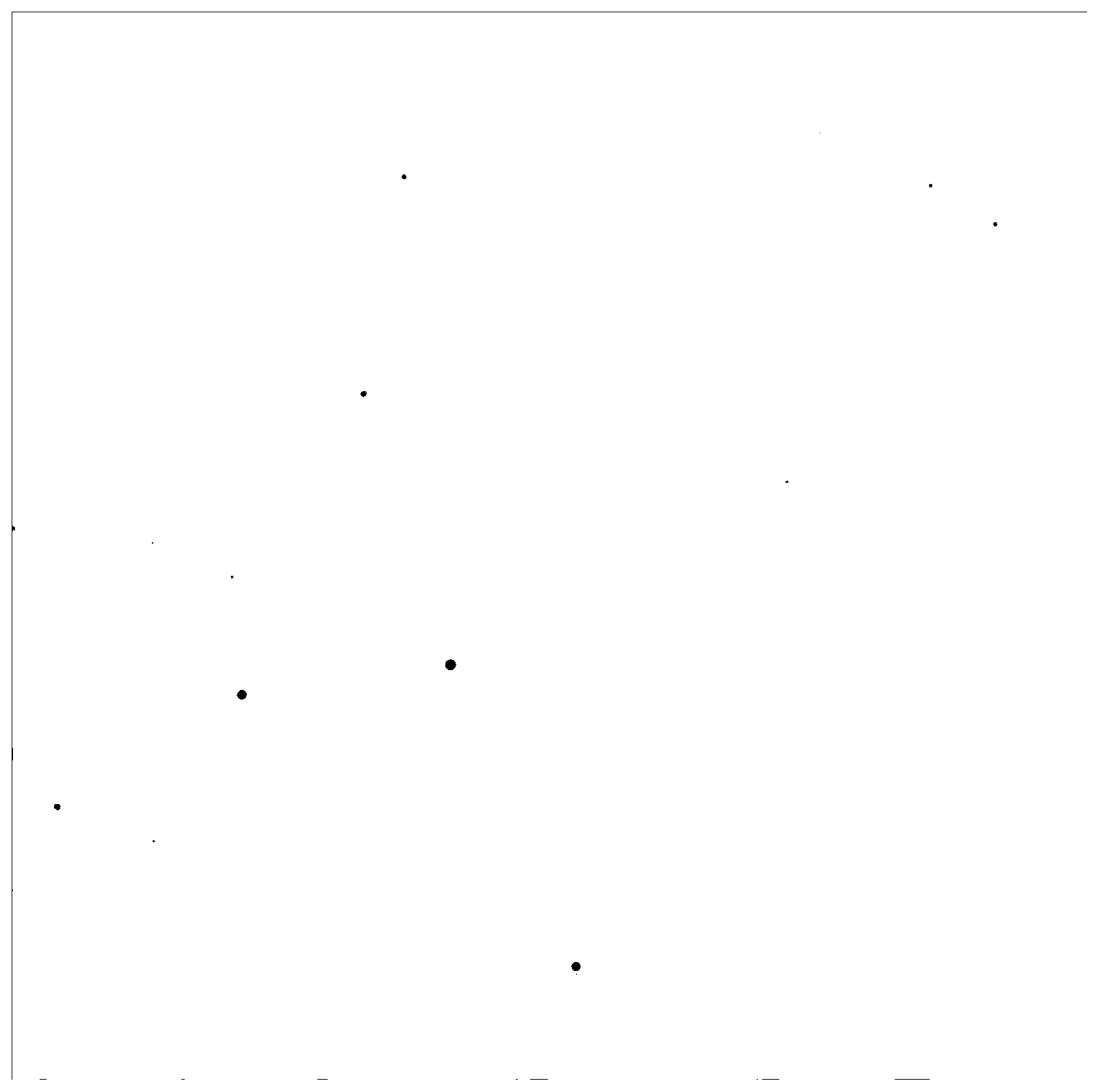}
  \caption{Example of the stack mask image for skycell
    skycell.1146.095 centered at ($\alpha,\delta$) = (11.934, -4.197)
    in the \rps{} filter, stack\_id 3956997.  The entire frame is
    largely unmasked after combining inputs, with the only remaining
    masks falling on the cores of bright stars, and in small regions
    around the brightest objects where the overlapping of diffraction
    spike masks have removed all inputs.}
  \label{fig:stack mask image}
\end{figure}

\begin{figure}
  \centering
  \includegraphics[width=0.9\hsize,angle=0,clip]{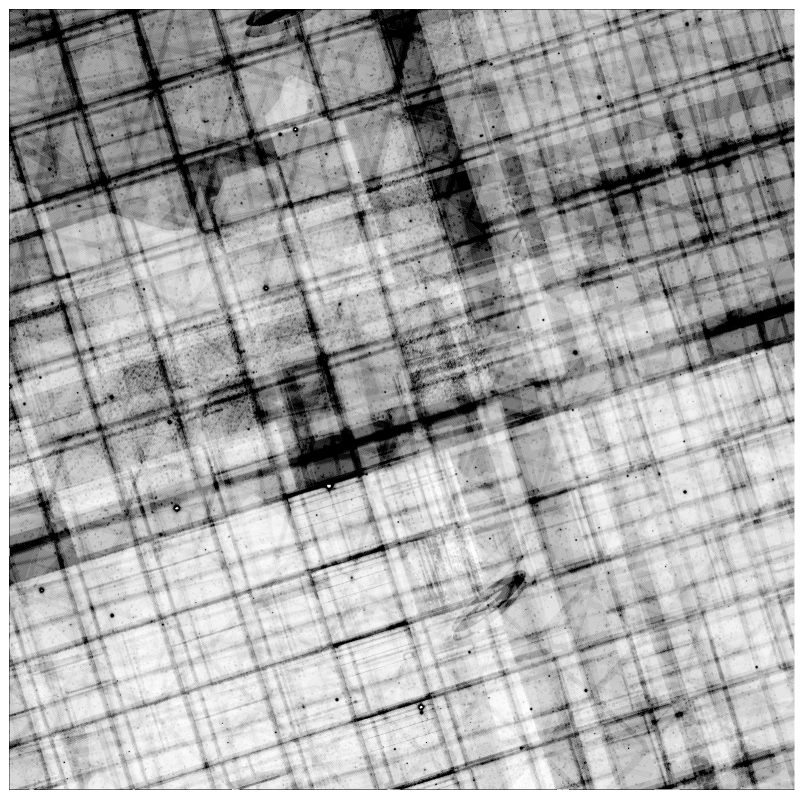}
  \caption{Example of the stack variance image for skycell 
    skycell.1146.095 centered at ($\alpha,\delta$) = (11.934, -4.197)
    in the \rps{} filter, stack\_id 3956997.  The variance
    map for this stack is reasonably smooth, with the mottled pattern
    from the inter-chip and inter-cell gaps printing through.  Some
    regions with higher variance are found where the number of inputs
    is lower.}

  \label{fig:stack wt image}
\end{figure}

\begin{figure}
  \centering
  \includegraphics[width=0.9\hsize,angle=0,clip]{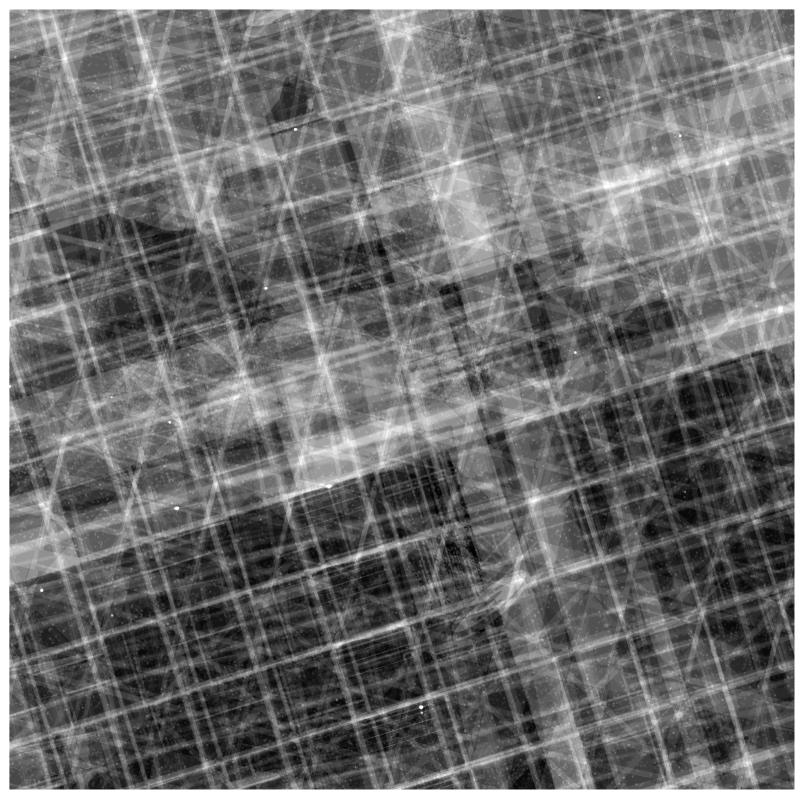}
  \caption{Example of the stack number image for skycell
    skycell.1146.095 centered at ($\alpha,\delta$) = (11.934, -4.197)
    in the \rps{} filter, stack\_id 3956997.  This map shows
    the number of inputs contributing to each pixel of the output
    stack.  Again, the pattern of the inter-chip and inter-cell gaps
    is visible, along with other mask features. }

  \label{fig:stack num image}
\end{figure}

\begin{figure}
  \centering
  \includegraphics[width=0.9\hsize,angle=0,clip]{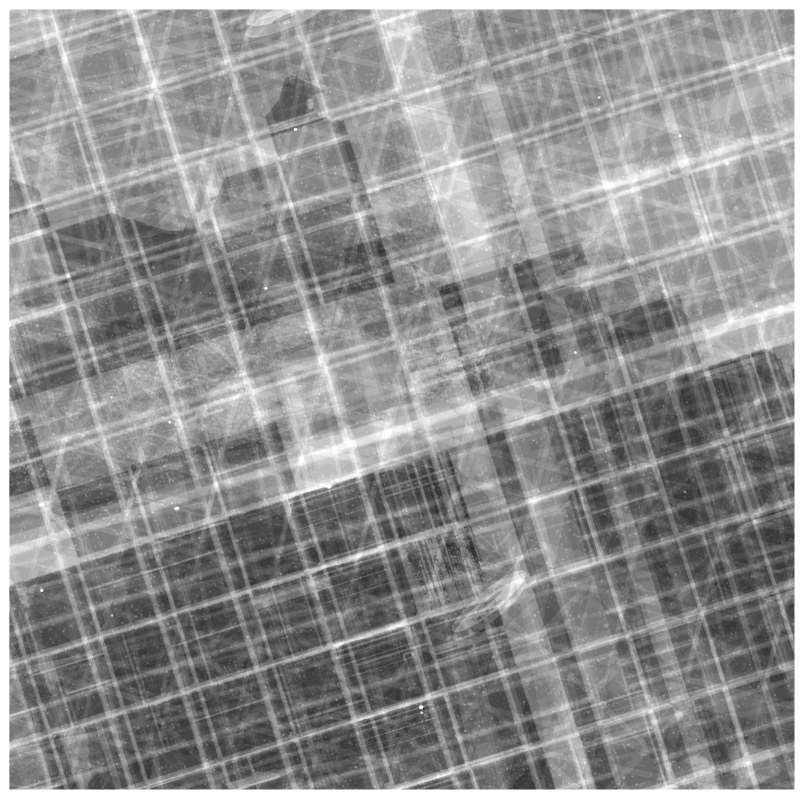}
  \caption{Example of the stack exposure time image for skycell
    skycell.1146.095 centered at ($\alpha,\delta$) = (11.934, -4.197)
    in the \rps{} filter, stack\_id 3956997.  Since the input
    exposures had exposures times of 40 and 60 seconds, the pattern
    observed here similar to, but subtly different from the number
    map.}
  \label{fig:stack exp image}
\end{figure}

\begin{figure}
  \centering
  \includegraphics[width=0.9\hsize,angle=0,clip]{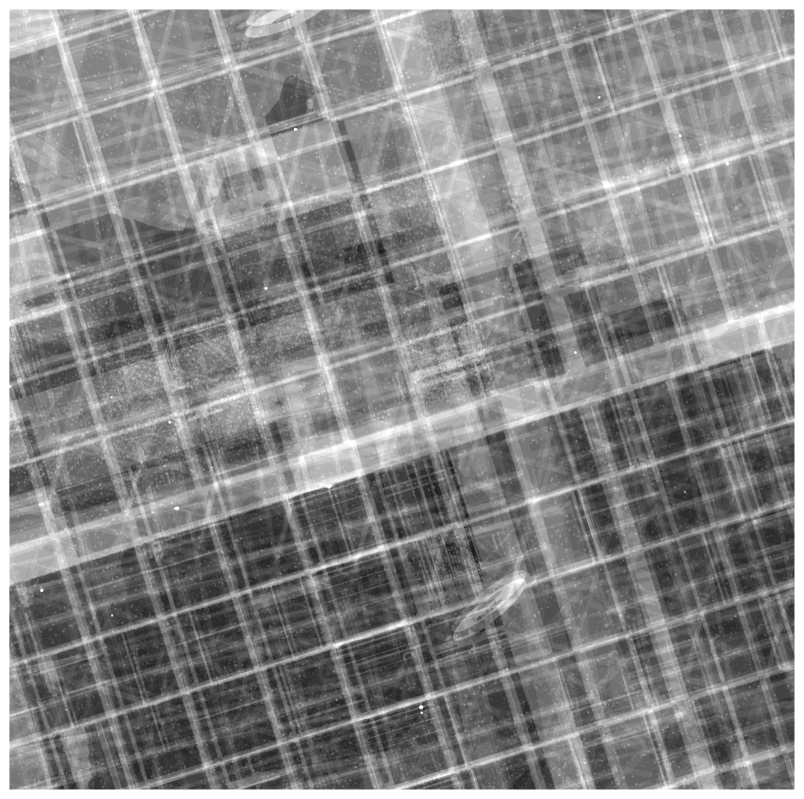}
  \caption{Example of the stack weighted exposure image for skycell
    skycell.1146.095 centered at ($\alpha,\delta$) = (11.934, -4.197)
    in the \rps{} filter, stack\_id 3956997.  This map shows
    the weighted average exposure time, as described in the text.  It
    is similar to the simple exposure time map, but shows how some
    input exposures have their contributions weighted down due to the
    observed larger image variances.}
  \label{fig:stack exp wtimage}
\end{figure}

Finally, a second pass at rejecting pixels is conducted, by extending
the current list to include pixels that are neighbors to many rejected
pixels.  The ISIS kernel used in the previous step is again used to
determine the largest square box that does not exceed the limit of
$0.25 \times \sum_{x,y} kernel^2$.  This square box is then convolved
with the rejected pixel mask to reject the neighboring pixels.  This
final list of rejected pixels is passed to the final combination,
which creates the final stack values from the weighted mean of the
non-rejected pixels.  Six total images are constructed for this final
stack: the image, its variance, a mask, a map of the exposure time per
pixel, that exposure time map weighted by the input image weight, and
a map of the number of inputs per pixel.  Examples of each output
image type for the stacking process are shown in
Figures~\ref{fig:stack image} through \ref{fig:stack exp wtimage}.

The convolved stack products are not retained, as the convolution is
used to ensure that the pixel rejection uses seeing-matched images.
This prevents any differences in the input PSF shape from skewing the
input pixel rejection.  We apply the normalizations and rejected pixel
maps generated from the convolved stack process to the original
unconvolved input images.  This produces an unconvolved stack that has
the optimum image quality possible from the input images.  Not
convolving does mean that the PSF shape changes across the image, as
the different PSF widths of the input images print through in the
different regions to which they have contributed.

%% Asinh compression

While IPP image products from single exposures use compressed 16-bit
integer images, this dynamic range is insufficient for the expected
scale of the stacked images.  This will lead either to truncation of
the extrema of the image, or quantized values that poorly sample the
image noise distribution.  Saving the images as 32-bit floating point
values would alleviate this quantization issue, at the cost of a large
increase in the disk space required for the stacked images.

Inspired by techniques used by SDSS \citep{1999AJ....118.1406L}, we
use the inverse hyperbolic sine function to transform the data.  The
domain of this function allows any input value to be converted.  In
addition, the quantization sampling can be tuned by placing the zero
of the inverse hyperbolic sine function at a value where the highest
sampling is desired.

Formally, prior to being written to disk, the pixel values are
transformed by $C = \alpha \asinh\left(\frac{L - \mathrm{BOFFSET}}{2.0
  \cdot \mathrm{BSOFTEN}}\right)$, where $L$ are the linear input
pixel values, $C$ the transformed values, and $\alpha = 2.5 \log_{10}(e)$.
BOFFSET centers the transformed values, and the mean of the linear
input pixel values is used.  BSOFTEN controls the stretch of the
transformation, and is set to $\sqrt{\alpha} \sigma_{L}$.  These
parameters are saved to the output image header.  The image is then
passed to the standard BSCALE and BZERO calculation and saved to disk.

To reverse this process (on subsequent reads of the image, for example
in warp-stack difference calculations), the BOFFSET and BSOFTEN
parameters are read from the header and the transformation inverted,
such that: $L = \mathrm{BOFFSET} + \mathrm{BSOFTEN} \cdot \left(\exp(C
/ \alpha) - \exp(-C / \alpha)\right)$.

\section{Difference Images}
\label{sec:diffs}

The image matching process used in constructing difference images is
essentially the same as for the stacking process.  An image is chosen
as a template, another image as the input, and after matching sources
to determine the scaling and transparency, convolution kernels are
defined that are used to convolve one or both of the images to a
target PSF.  The images are then subtracted, and as they should now
share a common PSF, static sources are largely subtracted (completely
in an ideal case), whereas sources that are not static between the two
images leave a significant remnant.  More information on the
difference image construction is contained in \citet{price2017}.  The
following section contains an overview of the difference image
construction used for the data in DR2.

The images used to construct difference images can be either
individual warp skycell frames or stacked images, with support for
either to be used as the template or input.  In general, for
differences using stacks, the deepest stack (or the only stack in the
case of a warp-stack difference) is used as the template.  The PV3
processing used warp-stack differences of all input warps against the
stack that was constructed from those inputs.  The same ISIS kernels
as were used in the stack image combination were again used to match
the stack PSF to the input warp PSF.  After convolution of the image
products, the difference is constructed for both the positive (warp
minus stack) and inverse (stack minus warp) to allow for the
photometry of the difference image to detect sources that both rise
and fall relative to the stack.  The convolution process grows the
mask fraction of pixels relative to the warp (the largest source of
masked pixels in these warp stack differences).  Any pixel that after
convolution has any contribution from a masked pixel is masked as
well, ensuring only fully unmasked pixels are used.

For warp-warp differences, such as those used for the ongoing Solar
System moving object search in nightly observations
\citep{2013PASP..125..357D}, the warp that was taken first is used as
the template.  As there is less certainty in which of the two input
images will have better seeing, a ``dual'' convolution method is used.
Both inputs are convolved to a target PSF that is not identical to
either input.  This intermediate target is essential for the case in
which the PSFs of the two inputs have been distorted in orthogonal
directions.  Simply convolving one to match the other would require
some degree of deconvolution along one axis.  As this convolution
method by necessity uses more free parameters, the ISIS kernels used
are chosen to be simpler than those used in the warp-stack
differences.  The ISIS widths are kept the same (1.5, 3.0, 6.0 pixel
FWHMs), but each Gaussian kernel is constrained to only use a
second-order polynomial.  As with the warp-stack differences, the mask
fraction grows between the input warp and the final difference image
due to the convolution.  For the warp-warp differences, each image
mask grows based on the appropriate convolution kernel, so the final
usable image area is highly dependent on ensuring that the telescope
pointings are as close to identical as possible.  The observing
strategy to enable this is discussed in more detail in
\citet{chambers2017}.

\section{Future Plans}
\label{sec:discussion}

Although the detrending and image combination algorithms work well to
produce consistent and calibrated images, having the PV3 processing of
the full $3\pi$ data set allows issues to be identified and solutions
created for future improvements to the IPP pipeline.  In addition, the
existence of the final calibrated catalog can be used to look for
issues that appear dependent on focal plane position.

One obvious way to make use of the PV3 catalog is to do a statistical
search for electronic crosstalk ghosts that do not match a known rule.
Given that bright stars do not equally populate all fields, choosing
exposures to examine to look for crosstalk rules is difficult.  The
current crosstalk rules were derived from expectations based on the
detector engineering, supplemented by rules identified largely based
on unmatched transients.  With the full catalog, identification of new
rules can be done statistically, looking at detection pairs that
appear more often than random.  

There is some evidence that we have not fully identified all of these
crosstalk rules, based on a study of PV3 images.  For example,
extremely bright stars may be able to create crosstalk ghosts between
the second cell column of OTA01 and OTA21, with possibly fainter
ghosts appearing on OTA11.  Despite the symmetry observed in the main
ghost rules, there do not appear to be clear examples of a similar
ghost between OTA47 and OTA66.  Examining this further based on the
PV3 catalog should provide a clear answer to this, as well as clarify
brightness limits below which the ghost does not appear.

The PV3 catalog may also allow better determination of which date
ranges we should use to build the dark model.  The date ranges
currently in use are based on limited sampling of exposures, and do
not have strong tests indicating that they are optimal.  By examining
the scatter between the detections on a given exposure and the catalog
average, we can attempt to look for increases in scatter that might
suggest that the dark model used is not completely correcting the
camera.  Looking at this based on the catalog would allow this
information to be generated without further image level processing.

In addition to improving the quality of the catalog for any future
reprocessing, there are a number of possible improvements that could
fix the image cosmetics.  A study of the burntool fits on stars that
have been badly saturated suggest that we may be able to improve the
trail fits by considering not the star center, but rather the edge of
saturation.  This restricts the fit to only consider the data along
the trail, and may improve the fit quality.  Implementing this change
would require additional bookkeeping of which pixels were saturated,
as the fits on subsequent exposures will need to skip these pixels
before fitting the persistence trail.  This is unlikely to seriously
impact the photometry of objects, but may improve the results of
stacks if fewer pixels need to be rejected.

The fringe model used currently is based on only a limited number of
days of data.  This means that the model calculated may not be fully
sensitive to the exact spectrum of the sky.  This may make the model
quality differ based on the date and local time of observation.  There
is some evidence that the fringe model does fit some dates better than
others, and so improving this by expanding the number of input
exposures may improve a wider range of dates.

Finally, a large number of issues arise due to the row-to-row bias
issues.  The PATTERN.ROW correction is used on a limited number of
cells, to minimize any possible distortion of bright stars or dense
fields by the fitting process.  As the row-to-row bias changes very
quickly in the y pixel axis and slowly along the x, it may be possible
to isolate and remove this signal in the Fourier domain.  Preliminary
investigations have shown that there is a small peak visible in the
power spectrum of a single cell, but determining the best way to
clip this peak to reduce the noise in the image space is not clear.

\section{Conclusion}

The Pan-STARRS1 PV3 processing has reduced an unprecedented volume of
image data, and has produced a catalog for the $3\pi$ Survey
containing hundreds of billions of individual measurements of
three billion astronomical objects.  Accurately calibrating
and detrending is essential to ensuring the quality of these results.
The detrending process detailed here produces consistent data, despite
the many individual detectors and their individual response functions.

From these individual exposures, we are able to construct images on
common projections and orientations, further removing the particulars
of any single exposure.  Furthermore, by created stacked images, we
can determine an estimate of the true static sky, providing a deep
data set that is ideal for use as a template for image differences.

The Pan-STARRS1 Surveys (PS1) have been made possible through
contributions by the Institute for Astronomy, the University of
Hawaii, the Pan-STARRS Project Office, the Max-Planck Society and its
participating institutes, the Max Planck Institute for Astronomy,
Heidelberg and the Max Planck Institute for Extraterrestrial Physics,
Garching, The Johns Hopkins University, Durham University, the
University of Edinburgh, the Queen's University Belfast, the
Harvard-Smithsonian Center for Astrophysics, the Las Cumbres
Observatory Global Telescope Network Incorporated, the National
Central University of Taiwan, the Space Telescope Science Institute,
and the National Aeronautics and Space Administration under Grant
No. NNX08AR22G issued through the Planetary Science Division of the
NASA Science Mission Directorate, the National Science Foundation
Grant No. AST-1238877, the University of Maryland, Eotvos Lorand
University (ELTE), and the Los Alamos National Laboratory.

\bibliographystyle{apj}
% \bibliography{lib}{}

\end{document}